\documentclass[fleqn,usenatbib]{mnras}

\usepackage{newtxtext,newtxmath}

\usepackage[T1]{fontenc}

\DeclareRobustCommand{\VAN}[3]{#2}
\let\VANthebibliography\thebibliography
\def\thebibliography{\DeclareRobustCommand{\VAN}[3]{##3}\VANthebibliography}

\usepackage{graphicx}	
\usepackage{amsmath}	
\usepackage{subfig}
\usepackage{empheq}

\title[Supernovae and AGN feedback]{Chemical evolution of elliptical galaxies I: supernovae and AGN feedback}

\author[M. Molero et al.]{
Marta Molero,$^{1,2}$\thanks{E-mail: marta.molero@phd.units.it}
Francesca Matteucci,$^{1,2,3}$
Luca Ciotti,$^{4}$
\\
$^{1}$Dipartimento di Fisica, Sezione di Astronomia, Università degli studi di Trieste, Via G.B. Tiepolo 11, I-34143 Trieste, Italy\\
$^{2}$INAF, Osservatorio Astronomico di Trieste, Via Tiepolo 11, I-34131 Trieste, Italy\\
$^{3}$INFN, Sezione di Trieste, Via Valerio 2, I.34127 Trieste, Italy\\
$^{4}$Dipartimento di Fisica e Astronomia, Università di Bologna, via Gobetti 93/2, I-40129 Bologna, Italy
}

\date{Accepted XXX. Received YYY; in original form ZZZ}

\pubyear{2015}

\begin{document}
\label{firstpage}
\pagerange{\pageref{firstpage}--\pageref{lastpage}}
\maketitle

\begin{abstract}
We study the formation and evolution of elliptical galaxies and how they suppress star formation and maintain it quenched. A one-zone chemical model which follows in detail the time evolution of gas mass and its chemical abundances during the active and passive evolution, is adopted. The model includes both gas infall and outflow as well as detailed stellar nucleosynthesis. Elliptical galaxies with different infall masses, following a down-sizing in star formation scenario, are considered. In the chemical evolution simulation we include a novel calculation of the feedback processes. We include heating by stellar wind, core-collapse SNe, Type Ia SNe (usually not highlighted in galaxy formation simulations) and AGN feedback. The AGN feedback is a novelty in this kind of models and is computed by considering a Bondi-Eddington limited accretion onto the central supermassive black hole. We successfully reproduce several observational features, such as the [$\alpha$/Fe] ratios increasing with galaxy mass, mass-metallicity, $\rm M_{BH}-\sigma$ and $\rm M_{BH}-M_{*}$ relations. Moreover, we show that stellar feedback and in particular Type Ia SNe, has a main role in maintaining quenched the star formation after the occurrence of the main galactic wind, especially in low-mass ellipticals. For larger systems, the contribution from AGN to thermal energy of gas appears to be necessary. However, the effect of the AGN on the development of the main galactic wind is negligible, unless an unreasonable high AGN efficiency or an extremely low stellar feedback are assumed. We emphasize the important role played by Type Ia SNe in the energy budget of early-type galaxies.

\end{abstract}




\section{Introduction}

Metal abundances are important since they are directly related to stellar mass loss and supernova (SN) ejecta, and they can provide constraints on the history of star formation (SF), initial mass function (IMF) and  metal enrichment history of the interstellar medium (ISM). 
Early-type galaxies (ETGs) are metal rich systems characterized by having high [$\alpha$/Fe] ratios in their dominant stellar population, with super-solar [Mg/Fe] in the nuclei of bright galaxies (\citealp{faber1992, carollo1993}). 
This is an important indicator of the fact that elliptical galaxies suffered a short duration of SF, since Type Ia SNe, which occur on a large interval of timescales, should not have had time to pollute significantly the ISM before the end of the SF, and therefore could not contribute to lower the [$\alpha$/Fe] ratio (according to the \textit{time-delay model}, \citealp{matteucci2001norecchi}). Moreover, the increase of the central [Mg/Fe] ratio with the stellar velocity dispersion suggests, always on basis of the time-delay model, that the more massive systems evolve faster than the less massive ones. This process is known as \textit{downsizing in star formation} (\citealp{1996cowie, 2004heavens, 2005treu}).

In order to account for this trend in the SF, the \textit{monolithic model} for the formation and evolution of ellipticals, first suggested by \citet{larson1975}, assumes that ellipticals suffer an intense SF and quickly produce galactic winds when the energy injected into the ISM equates the potential energy of the gas. SF is then quenched and galaxies are evolving passively afterwards. Then, in order to reproduce the increasing trend of the [Mg/Fe] with galactic mass, \citet{matteucci1994} first computed models for ellipticals with a shorter period of SF in larger systems, assuming an increasing efficiency of SF with the galactic mass. As a consequence, galactic winds occur earlier in more massive galaxies (\textit{inverse wind scenario}) and the [$\alpha$/Fe] ratio increases with galaxy stellar mass.

Galactic outflows are both theoretically expected (\citealp{1988tomisaka}) and observationally detected (\citealp{1990heckman}). Physically, they are expected to be driven by the energy released from stars and SNe (\citealp{1985chevalier}), as well as from supermassive black holes (SMBH, \citealp{begelman1991}). Active galactic nucleus (AGN) feedback is fundamental to control the BH growth and the AGN activity itself, by regulating the evolution of the physical properties of the surrounding gas, and therefore the BH accretion and luminosity. Outflows and feedback are fundamental aspects of galaxy formation and evolution, however the underlying physical mechanisms are complex and it is still debated whether AGN feedback is the main driver of galaxy evolution and to what level it impacts on the physical properties of the bulk of the gas in galaxies (\citealp{2021valentini}). Indeed, much more investigation is still needed. 
There exist many theoretical works which address this important question by adopting different treatments of the feedback processes: the energetic output is usually parametrized by invoking stellar winds, SNe and/or AGNs, or a combination of these.  

Results of semi-analytic works, which have been conducted in recent decades, point out that the most important mechanism able to suppress the SF activity is stellar feedback (both in form of stellar winds and SN explosions), at least in relatively low-mass elliptical galaxies (\citealp{1999somerville, 2003benson, 2006bower, 2012bower, 2017pan}). For high mass galaxies instead, AGN feedback can efficiently regulate the SF activity (\citealp{2012silk, 2018li}).
On the other hand, theoretical studies based on large-scale cosmological simulations find that the energy feedback from Type II SNe alone is not enough to quench the SF activity both in low- and in high-mass elliptical galaxies and an energy source from radiation, wind and radio jets from the central AGN is needed (\citealp{2006croton, 2007lu, 2015choi, 2016dave, 2017taylo, 2017weinberger}). These results are often a consequence of a poor modelling of the energy feedback from SNe, especially from Type Ia SNe, being one of the most difficult processes to model in galaxy-formation simulations (see \citealp{2003kawata} and references therein, but see also \citealp{2006scannapieco, 2008scannapieco, 2015jime}).

However, as also pointed out by \citet{2018li}, the cosmological simulations, compared to analytical models, are better at capturing the environmental effects occurring during the cosmological evolution of galaxies, but the scales on which the feedback processes operate are much smaller that the typical resolution of the simulations and a much higher resolution is needed in order to focus on the relatively small scales of influence of the different feedback processes (even if the situation is improving in recent years: e.g. \citealp{2016curtis, 2020Costa, 2021alcazar}). 

Many hydrodynamical simulations have been carried out in this direction in order to study the feedback processes in detail, both focusing on the effect of the AGN feedback (\citealp{1995binney, 2012choi, 2014gan, 2017ciotti, 2018yuan, 2022ciotti}) and on the role of SN feedback (\citealp{1991ciotti, 2018smith, 2021lanfranchi}). The main advantage of hydrodynamical models is that complex physical effects can be taken into account with high accuracy. 
However, the computational times of those simulations are long and sometimes it is useful to search for less time-consuming solutions, usually represented by one-zone models. In this work, we adopt a one-zone chemical evolution model. These models are very detailed in computing the chemical abundances and can take into account dynamical processes in a simple way (\citealp{2005sazonov, 2008ballero, 2008matteucci, 2011Lusso}). Interesting cases that can be identify by those models can then be simulated in much more detail with hydrodynamical codes.

In this work, we adopt an updated version of \citet{matteucci1994} chemical evolution model for elliptical galaxies, where the SN rates are computed in details as well as the stellar nucleosynthesis. The main novelty of this model is the inclusion of the AGN feedback, besides that of SNe and stellar winds. In particular,  we study the evolution of ETGs with different initial infall of gas mass (between $\rm 10^{10}\ -\ 5\times10^{12}\ M_\odot$). The evolutionary scenario that we consider is the following: ellipticals are formed by infall of gas in a primordial dark matter halo and its evolution is influenced by infall and outflow of gas as well as by stellar nucleosynthesis. The system goes through an early intense burst of SF, which is then quenched when strong galactic winds are produced and the galaxy evolves passively afterwards. This happens when the thermal energy of the gas in the ISM exceeds its binding energy. We study both the case in which the gas is thermalized only by stellar winds and SNe of all types, with particular attention to Type Ia SNe, and the case in which AGN feedback also contributes to the thermal energy of the gas. 

The paper is organized as follow. In Section 2 we present the chemical evolution simulations with the basic equations which are used to describe the evolution of the gas mass, the nucleosynthesis prescriptions and the energetic treatment. Then, in Section 3 we show the results obtained when no AGN feedback is adopted. Section 4 is devoted to the description of the adopted treatment for the BH accretion, luminosity and feedback and also results obtained with the new energy formulation are presented. Finally, in Section 5 we present our discussion and conclusions.



\section{CHEMICAL EVOLUTION MODEL}

In order to study the chemical evolution of ETGs, we adopt a new model based on the main assumptions presented in \citet{matteucci1994} and similar to the most recent model of \citet{demasi2018}. 
The model is one zone but it can be easily extended to be multi-zone. It assumes instantaneous and complete mixing of gas. It is able to follow in detail the evolution of 22 chemical species, from H to Eu, from the beginning of SF up to the present time. It is assumed that galaxies form by infall of primordial gas in a pre-existing diffuse dark matter halo with a mass about 10 times the total mass of the galaxy. Stellar lifetimes are taken into account, thus relaxing the instantaneous recycling approximation (IRA). An early intense burst of star formation is followed by a massive galactic wind. After this main wind the galaxy can continue to loose mass or just stop the wind, depending on the assumptions made on feedback and gravitational potential, as we will describe in the next Sections.

\subsection{Basic equations}

The fundamental equations which describe the temporal evolution of the mass fraction of the generic element $i$ in the gas $\rm G_i(t)$ have the following form (for details, see \citealp{matteucci2012}):
\begin{align*}
    \dot{G}_i(t) &= - \psi(t)X_i(t) + X_{i,inf}(t)\dot{G}_{i,inf}(t) - X_i(t)\dot G_{i,w}(t)+\\
    &+ \dot{R}_i(t) - X_i(t)\dot{M}_{BH}(t),
\end{align*}
where $\rm X_i(t)$ is the abundance by mass of the element $i$ at the time $t$ ($\rm \sum_{i}X_i=1$) and $\rm X_{i,inf}(t)$ is the abundance of the element $i$ of the infalling gas. The terms of the right-hand side of the equation are:

\begin{itemize}
    \item The first term represents the rate at which chemical elements are subtracted by the ISM to be included in stars. $\rm \psi(t)$ is the star formation rate (SFR), which represents how many solar masses of gas are turned into stars per unit time. The SF is assumed to stop as soon as galactic winds are generated. Until that moment, the SF follows a Schmidt-Kennicutt law with $\rm k=1$ (\citealp{schmidt, kennicutt}), so that:
    \begin{equation}
    \psi(t)= 
    \begin{cases}
    \ \nu G(t)^k\ \ if\ \ t < t_{GW} \\
    \ 0\ \ \ \ \ \ \ \ \ \ \ if\ \ t \ge t_{GW} 
    \end{cases}
\end{equation}
    where the constant $\rm \nu$ is the star formation efficiency expressed in Gyr$^{-1}$ and represents the inverse of the time needed to convert all the gas into stars. We assume $\rm \nu$ to increase with the galactic mass in order to reproduce the so called \textit{inverse wind model} (\citealp{matteucci1994, matteucci1998}). 
    
    \item The second term is the rate at which the chemical elements are accreted through infall of gas. It is given by the following relation:
    \begin{equation}
        \dot G_{i,inf} \propto X_{i,inf}e^{-t/\tau_{inf}},
    \end{equation}
    where $\mathrm{X_{i,inf}}$ represents the chemical abundance of the element $i$ of the infalling gas (here assumed to be primordial and therefore with no metals) and $\rm \tau_{inf}$ is the infall timescale, defined as the time at which half of the total mass of the galaxy has been assembled. The reason for the choice of a continuous infall rather than hierarchical mergers to form ellipticals is due to the fact that mergers rise some important problems in reproducing the properties of stellar populations in these galaxies. In particular, in \citet{2008Pipino} it was explored the effect of dry mergers on the chemical properties of stars in elliptical galaxies. It was found that a series of multiple dry mergers (with no star formation in connection with the mergers), involving building blocks that have been created ad hoc to satisfy the [Mg/Fe]-mass relation observed in these galaxies, cannot fit the mass metallicity relation and vice versa. 
    In conclusion, dry mergers alone seem not to explain the need of a more efficient star formation in the more massive galaxies, as suggested by the [Mg/Fe]-mass relation, as well as the late-time assembly suggested in the hierarchical paradigm to recover the galaxy downsizing. In addition, there are also simulations taking into account cosmological infall. In particular, in \citet{2008colavitti}, a cosmological infall law is derived based on dark matter halo properties and this resembles the exponential infall law predicted for the Galaxy (\citealp{1997chiappini}). Therefore, we think that a continuous gas infall is more appropriate to reproduce the chemical properties of ellipticals, as we will see in the next paragraphs.
    \item The third term represents the outflow rate of the element $i$ due to galactic winds developing when the thermal energy of the gas exceeds its binding energy (see Section \ref{sec: energetics}). The outflow rate has the following law:
    \begin{equation}
    \dot G_{i,w}(t)= 
    \begin{cases}
    \ 0\ \ \ \ \ \ \ \ \ \ \ \ if\ \ t < t_{GW} \\
    \ \omega_i G(t)\ \ if\ \ t \ge t_{GW} 
    \end{cases}
    \end{equation}
    where $\omega_i$ is the wind parameter (the so called \textit{mass-loading} factor) for the element $i$. It is a free adimensional parameter tuned to reproduce specific observational features of the simulated galaxy. Here we do not adopt differential wind, so the mass-loading factor is the same for all the chemical elements.
    
    \item The fourth term $\mathrm{R_i(t)}$ represents the fraction of matter which is returned by stars into the ISM through stellar winds, SN explosions and merging neutron stars (MNS), in the form of the element $i$. Namely, it represents the rate at which each chemical element is restored into the ISM by all stars dying at the time $t$. $\mathrm{R_i(t)}$ depends on the initial mass function (IMF, $\mathrm{\phi(m)}$), whose different parametrizations adopted will be described in Section \ref{sec: mass-metallicity relations}.
    \item Finally, the last term is the rate at which the mass fraction of gas in the form of the chemical element $i$ is accreted by the BH. Details of this term will be further described in Section \ref{sec: black hole accretion and agn feedback}.
    For Type Ia SNe we assumed a single-degenerate scenario in which SNe arise from the explosion via C-deflagration of a C-O white dwarf in a close binary system as it reaches the Chandrasekhar mass due to accretion from its red giant companion. Then, following \citet{matteuccigreggio1986, matteucci2001norecchi}, the rate is given by:
    \begin{equation}
        R_{SNIa}=A\int_{M_{B_m}}^{M_{B_M}}dM_B\phi(M_B)\int_{\mu_m}^{0.    5}f(\mu)\psi(t-\tau_m)d\mu,
    \end{equation}
    where $\mathrm{M_B}$ is the mass of the whole binary system and     $\mathrm{M_{B_m}}$ and $\mathrm{M_{B_M}}$ are the minimum and     maximum mass of the progenitor systems equal to $\mathrm{3\     M_\odot}$ and $\mathrm{16\ M_\odot}$, respectively. The parameter     $\mathrm{\mu=M_2/M_B}$ is the mass fraction of the secondary     component of the binary system, which follows the distribution:
    \begin{equation}
        f(\mu)=2^{\gamma+1}(\gamma+1)\mu^\gamma
    \end{equation}
    with $\mathrm{\gamma=2}$. Finally, $A$ represents the fraction of binary systems which are able to give rise to a Type Ia SN explosion. It is a free parameter, constrained in order to     reproduce the present-day observed Type Ia SNe rate of \citet{cappellaro1999}.
    
    For Type II SNe, their rate is computed as:
    \begin{equation}
    \begin{split}
        R_{SNII} & =(1-A_B)\int_{M_{up}}^{M_{B_M}}\psi(t-\tau_m)\phi(m)dm +\\
        & +\int_{M_{B_M}}^{M_{WR}}\psi(t-\tau_m)\phi(m)dm +\\
        & +\int_{M_{WR}}^{M_{max}} \psi(t-\tau_m)\phi(m)dm +\\
        & +\alpha_{Ib/c}\int_{12}^{20} \psi(t-\tau_m)\phi(m)dm,
    \end{split}
    \end{equation}
    \\
    where the lower extreme of the first integral is $\mathrm{M_{up}}$, namely the limiting mass for the formation of a degenerate C-O core, defined in the range $\mathrm{6-8\ M_\odot}$. The mass $\mathrm{M_{WR}}$ is the limiting mass for the formation of a Wolf-Rayet star. Above this mass, single stars become Wolf-Rayet and explode as Type Ib/c SNe. Their rates are represented by the last two integrals. In fact, Type Ib/c supernovae can originate either from the explosion of single Wolf-Rayet stars with masses $\mathrm{\ge\ M_{WR}}$, or from massive binary systems made of stars with masses in the range $\mathrm{12\le M/M_\odot \le 20}$. $\mathrm{M_{max}}$ is the maximum mass assumed for existing stars and it can be as high as $\mathrm{100\ M_\odot}$. Finally, $\mathrm{\alpha_{Ib/c}}$ represents the fraction of massive binary systems in the range $\mathrm{12\le M/M_\odot \le 20}$ which can give rise to SNe Ib/c.
    
\end{itemize}

\subsection{Nucleosynthesis prescriptions}

For all the stars sufficiently massive to die in a Hubble time, the following stellar yields have been adopted:

\begin{itemize}
    \item For low and intermediate mass stars (LIMS in the $\rm \mathrm{0.8-8}\ M_\odot$ range) we include the metallicity-dependent yields of \citet{vandenhoek}.
    \item For massive stars we assume yields of \citet{francois2004}.
    \item For Type Ia SNe we include yields of \citet{iwamoto1999}.
    \item For r-process elements we adopted the best models of \citet{molero22021}: r-process elements are produced by both MNS (with a yield of $\rm 3\times10^{-6}\ M_{\odot}$ per merging event) and by magneto-rotational driven SNe (MRD-SNe), with a yield equal to that of the theoretical calculations of \citealp{nishimura2017}, their model L0.75. In particular, we assume that only 1\% of the stars with initial mass in the $\rm 10-80\ M_\odot$ range would explode as MRD-SNe.  
\end{itemize}

\subsection{Energy prescriptions}
\label{sec: energetics}

The existence of a wind phase at some stage of evolution of elliptical galaxies is required in order to both  explain the observed iron abundance in the intracluster medium and  avoid overproducing gas. Galactic winds develop when the thermal energy of the gas, $\mathrm{E_{gas}^{th}(t)}$, exceeds its binding energy $\mathrm{E_{gas}^b(t)}$ (see \citealp{matteucci1994, bradamante1998}):
\begin{equation}
    E_{\mathrm{gas}}^{\mathrm{th}}(t) \geq E_{\mathrm{gas}}^\mathrm{b}(t).
\label{eq: gw}
\end{equation}

In the next Sections we will focus on the description of the different contributions to those two terms.

\subsubsection{Gas thermal energy}
\label{sec: thermal energy}
The gas thermal energy is given by the sum of the thermal energy deposited in the gas by SN explosions, $\mathrm{E_{SN}^{th}(t)}$, stellar winds $\mathrm{E_{wind}^{th}(t)}$ and  AGN feedback $\rm E_{AGN}^{th}(t)$:
\begin{equation}
    E_{gas}^{th}(t)=E_{SN}^{th}(t)+E_{wind}^{th}(t)+E_{AGN}^{th}(t).
    \label{eq: thermal_energy}
\end{equation}

In this Section we will focus on the contribution by SNe and stellar winds. The AGN feedback is further described in Section \ref{sec: black hole accretion and agn feedback}.

In particular, $\rm {E_{SN}^{th}(t)}$ is given by the contribution of both Type II SNe ($\rm E_{II}^{th}$, here Type Ib/c SNe are included in the Type II SNe) and Type Ia SNe ($\rm E_{Ia}^{th}$), while $\rm {E_{wind}^{th}(t)}$ is given by the contribution of both stellar winds from massive stars ($\rm E_{W}^{th}$) and winds from LIMS ($\rm E_{\sigma}^{th}$). So that:
\begin{equation}
\begin{split}
    E_{SN}^{th}(t)=E_{II}^{th}(t)+E_{Ia}^{th}(t)\\
    E_{wind}^{th}(t)=E_{W}^{th}(t)+E_{\sigma}^{th}(t).
\end{split}
\end{equation}
\\
We have:
\begin{equation}
\begin{split}
    E_{II}^{th}(t)=\int_{0}^{t}\epsilon_{II}R_{II}(t')dt'\\
    E_{Ia}^{th}(t)=\int_{0}^{t}\epsilon_{Ia}R_{Ia}(t')dt'\\
    E_{W}^{th}(t)=\int_{0}^{t} \int_{8}^{m_{up}}\phi(m)\psi(t')\epsilon_{W}dmdt'\\
    E_{\sigma}^{th}(t)=\int_{0}^{t}\int_{0.8}^{8}\phi(m)\psi(t')\sigma^2(t') dmdt',
\end{split}
\end{equation}
with $\sigma^2=0.335GM_*(t)/R_e$ being the stellar velocity dispersion. $\mathrm{R_{II}}$ and $\mathrm{R_{Ia}}$ are the rates of Type II and Type Ia SNe, respectively that we showed in the previous section and the terms $\mathrm{\epsilon_{II/Ia}}$ and $\mathrm{\epsilon_{w}}$ are the energies injected into the ISM from supernova explosions and stellar winds from massive stars, respectively. In particular:
\begin{equation}
\begin{split}
    \epsilon_{II}=\eta_{II}E_0\\
    \epsilon_{Ia}=\eta_{Ia}E_0\\
    \epsilon_{W}=\eta_{W}E_{W},
\end{split}
\end{equation}
where $\mathrm{E_0=10^{51}\ erg}$ is the total energy released by a supernova explosion and $\mathrm{E_{wind}=10^{49}\ erg}$ is the energy injected into the ISM by a typical massive star during its all lifetime. $\mathrm{\eta_{II}}$, $\mathrm{\eta_{Ia}}$ and $\mathrm{\eta_{W}}$ are the efficiencies of energy transfer from supernova Type II, Type Ia and stellar winds into the ISM, respectively. According to \citet{cioffi1988}, due to significant cooling by metal ions, only a few per cent of the initial $\mathrm{10^{51}\ erg}$ can be provided to the ISM by Type II SNe. On the other hand, since Type Ia SNe explosions occur in a medium already heated by Type II SNe, they can contribute with a  higher percentage of their energy budget (\citealp{recchi2001, matteucci2001, pipino2002, demasi2018}). In this work we assumed an efficiency of $3 \%$ for Type II SNe and stellar winds (see \citealp{bradamante1998, 2004melioli}) and tested three different values for Type Ia SNe: $80 \%$, $30 \%$ and $10 \%$, simulating different cooling conditions. 

\subsubsection{Gas binding energy}

Following \citet{1991bertin}, elliptical galaxies have their luminous mass embedded in massive and diffuse dark matter halos. In this context, the binding energy of the gas can be expressed as:
\begin{equation}
    E_{gas}^b(t)=W_L(t)+W_{LD}(t),
\end{equation}
where $\mathrm{W_L(t)}$ is the gravitational energy of the gas due to the luminous matter, given by
\begin{equation}
    W_L(t)=-q_LG\frac{M_{gas}(t)M_{L}(t)}{R_e},
\end{equation}
with $\rm M_{L}(t)$ being the total baryonic mass at the time t, $\mathrm{R_e}$ the effective radius and $\mathrm{q_L}=1/2$. $\mathrm{W_{LD}(t)}$ is the gravitational energy of the gas due to the interaction of luminous and dark matter:
\begin{equation}
    W_{LD}(t)=-\Tilde{\omega}_{LD}G\frac{M_{gas}(t)M_{DM}}{R_e},
\end{equation}
where $\mathrm{M_{DM}}$ is the mass of the dark matter halo and
\begin{equation}
    \Tilde{\omega}_{LD}=\frac{1}{2\pi}\frac{R_e}{R_{DM}}\Big[1+1.37\Big(\frac{R_e}{R_{DM}}\Big)\Big]
\end{equation}
is the interaction term, with $\mathrm{R_{DM}}$ being the radius of the dark matter halo. According to \citet{1991bertin}, the relations for the gravitational interaction between the gas mass and the total luminous mass of the galaxy, and between the gas mass and the dark matter, are valid for $\mathrm{R_e/R_{DM}}$ defined in the range $\mathrm{0.10 - 0.45}$, at least for massive elliptical galaxies. Here we adopted  $\mathrm{R_e/R_{DM}=0.1}$, since that was considered being the best value in previous works (e.g.: \citealp{1992matteucci,demasi2018}).

\subsubsection{Galaxy binding energy}

The binding energy of the galaxy is given by:
\begin{equation}
    E_{gal}^b(t)=B_L(t)+B_{LD}(t),
\end{equation}
where $\rm B_L(t)$ is the gravitational energy of the galaxy due to the luminous matter, given by
\begin{equation}
    B_L(t)=-q_LG\frac{M^2_L(t)}{R_e}
\end{equation}
and $\rm B_{LD}(t)$ is the gravitational energy of the galaxy due to the interaction between luminous and dark matter:
\begin{equation}
    B_{LD}(t)=-\Tilde{\omega}_{LD}G\frac{M_{L}(t)M_{DM}}{R_e}.
\end{equation}

\section{RESULTS}
\label{sec: models}

The model for the chemical evolution of elliptical that we run is similar to one of the best models reported by \citet{demasi2018} (their model 02b), who used a chemical evolution code similar to the one used here. 

The model explores the evolution of elliptical galaxies in the baryonic mass range $\mathrm{10^{10}-5\times10^{12}\ M_\odot}$. The effective radius $\mathrm{R_{e}}$ increases with the baryonic mass and, according to the \textit{inverse wind scenario} (\citealp{matteucci1994}), the star formation efficiency $\mathrm{\nu}$ increases as well while the infall timescale $\mathrm{\tau}$ decreases.

In Table \ref{tab: models}, we report the adopted parameters. In particular, in the $\mathrm{1^{st}}$, $\mathrm{2^{nd}}$, $\mathrm{3^{rd}}$ and $\mathrm{4^{th}}$ columns we report the adopted infall mass, the star formation efficiency, the infall timescale and the effective radius, respectively. In the $\mathrm{5^{th}}$, $\mathrm{6^{th}}$ and $\mathrm{7^{th}}$ columns we report the predicted final stellar mass, time of the onset of the galactic wind and present day Type Ia Sne rate. In the last column we report the adopted IMF. In particular, according to \citet{demasi2018}, models with constant IMF for galaxies of different mass fail in reproducing the observed trends with galactic mass. They tested a varying IMF and found a better agreement with data by assuming that the IMF goes from being bottom heavy in less massive galaxies to top heavy in more massive ones, producing a downsizing in star formation, favoring massive stars in larger galaxies. The adopted IMF in this work are:

\begin{itemize}
    \item A \citet{scalo1986S} IMF for low mass galaxies:
    \begin{equation}
    \phi(m) \propto 
    \begin{cases}
    \ m^{-2.35}\ \ \mathrm{if}\ \ 0.1\leq m/M_\odot < 6\\
    \ m^{-2.7}\ \ \ \ \mathrm{if}\ \ 6 \leq m/M_\odot \leq 100.
    \end{cases}
    \end{equation}
    \item A \citet{salpeter1955} IMF for intermediate mass galaxies:
    \begin{equation}
        \phi(m)\propto m^{-2.35}.
    \end{equation}
    \item A \citet{arimoto1987} IMF for high mass galaxies:
    \begin{equation}
        \phi(m)\propto m^{-1.95}.
    \end{equation}
\end{itemize} 

As a first step, we try to reproduce the main chemical properties of the stellar populations dominating the spectra of ETGs.

\begin{table*}
\centering
\hspace{-0.5 cm}
\caption{\label{tab: models} Parameters of the model described in section \ref{sec: models}. We adopted different values for the star formation efficiency $\nu$, the infall timescale $\tau$ and the effective radius $\mathrm{R_{eff}}$ (columns 2, 3 and 4, respectively) for the different infall masses $\mathrm{M_{i}}$ (column 1). In column 5, 6 and 7 we report the predicted final stellar mass $\mathrm{M_{*_f}}$, the predicted time for the onset of the galactic wind $\mathrm{t_w}$ and the predicted rate of Type Ia SNe $\mathrm{R_{Ia}}$. Finally, on the last column we specify the adopted IMF.}
\begin{tabular}{cccccccc}
\hline
  $\mathrm{M_i}$ $(\mathrm{M_{\odot}})$ & $\mathrm{\nu}$ $(\mathrm{Gyr^{-1}})$ & $\mathrm{\tau_i}$ $\mathrm{(Gyr)}$ & $\mathrm{R_{e}}$ $\mathrm{(kpc)}$ & $\mathrm{M_{*_f}}$ $\mathrm{(M_\odot)}$ & $\mathrm{t_w}$ $\mathrm{(Gyr)}$ & $\mathrm{R_{Ia}}$ $\mathrm{(SN/century)}$ & $\mathrm{IMF}$\\
\hline
   $1\times10^{10}$ & $3.0$ & $0.5$ & $1$ & $1.0\times10^{9}$ & $0.37$ & $0.004$ & Scalo\\
   $5\times10^{10}$ & $6.0$ & $0.4$ & $2$ & $1.5\times10^{10}$ & $0.35$ & $0.031$ & Salpeter\\
  $1\times10^{11}$ & $10$ & $0.4$ & $3$ & $2.0\times10^{10}$ & $0.33$ & $0.072$ & Salpeter\\
  $5\times10^{11}$ & $15$ & $0.3$ & $6$ & $1.5\times10^{11}$ & $0.33$ & $0.524$ & Arimoto\&Yoshii\\
  $1\times10^{12}$ & $22$ & $0.2$ & $10$ & $2.0\times10^{11}$ & $0.25$ & $1.178$ & Arimoto\&Yoshii\\
  $5\times10^{12}$ & $60$ & $0.1$ & $12$ & $1.5\times10^{12}$ & $0.19$ & $5.024$ & Arimoto\&Yoshii\\
\hline
\end{tabular}%
\end{table*}



\begin{figure*}
\begin{center}
 \subfloat{\includegraphics[width=1\columnwidth]{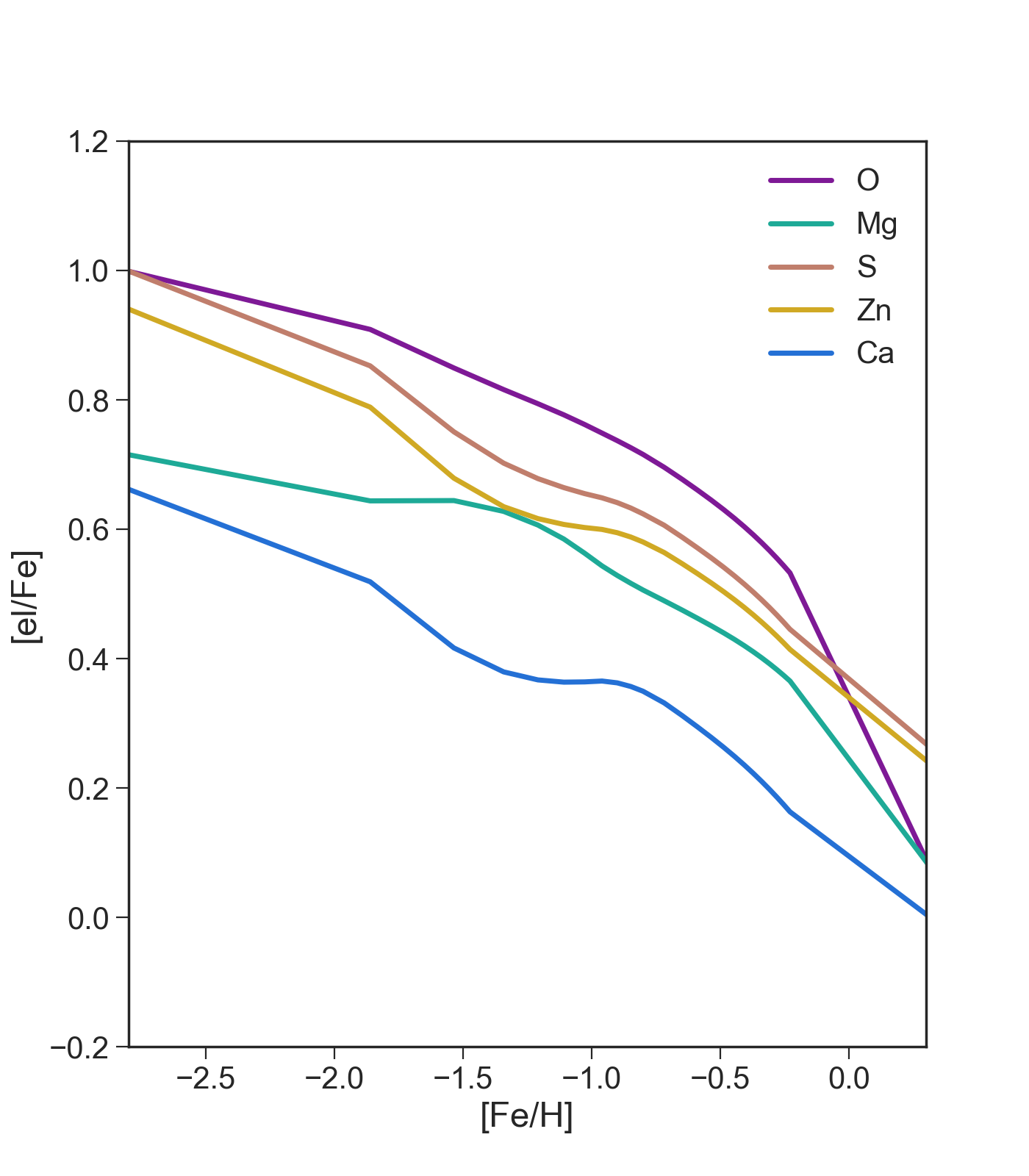}}
 \hfill
 \subfloat{\includegraphics[width=1\columnwidth]{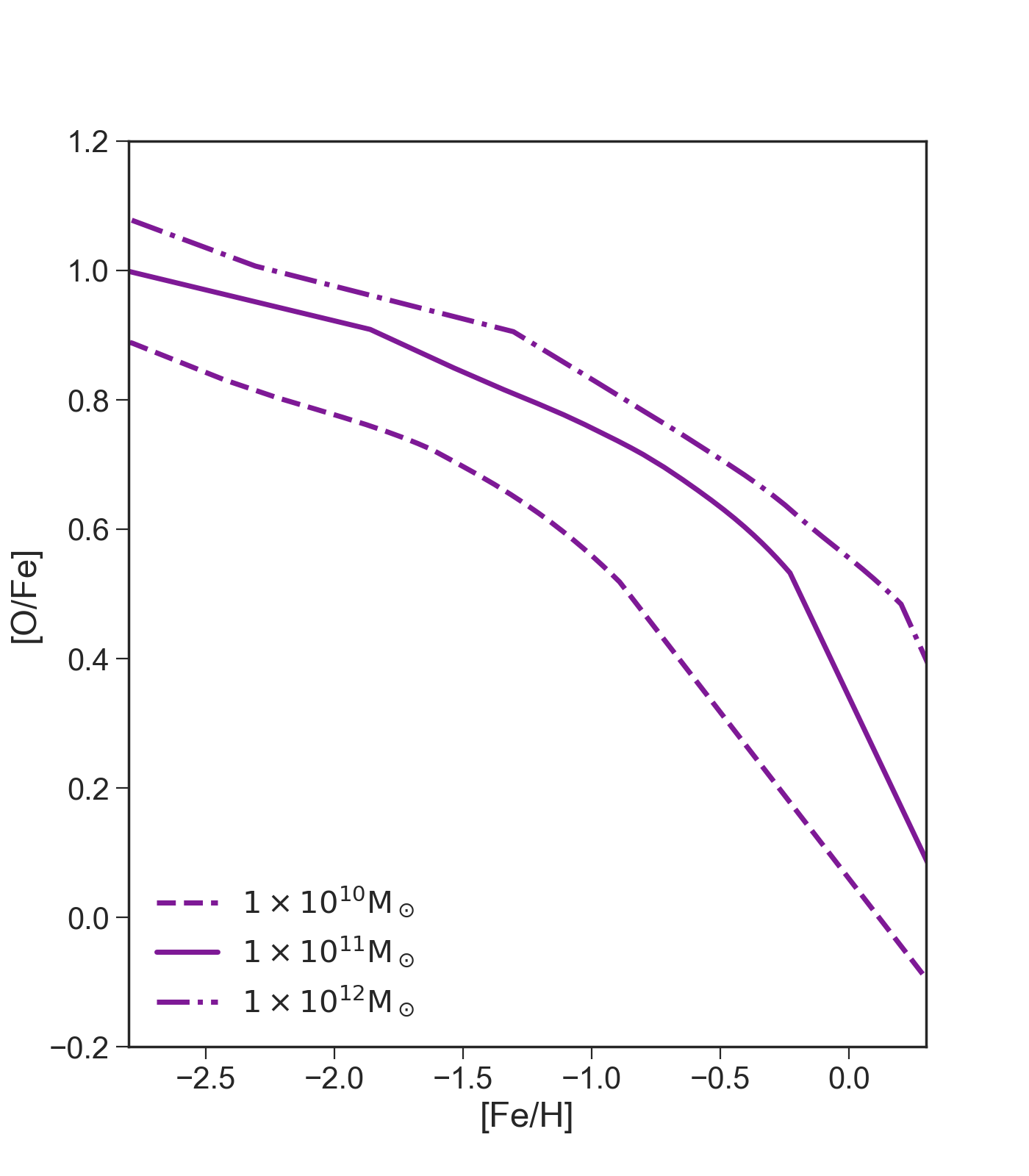}}%
 \caption{Left panel: Predicted abundances ratios in the ISM as functions of [Fe/H] for O, Mg, S, Zn and Ca for an elliptical galaxy with infall gas mass of $\rm M_i=10^{11} M_\odot$. Right panel: Predicted [O/Fe] abundance ratio in the ISM as a function of [Fe/H] for galaxies with $\rm 10^{10} M_\odot$ (dashed line), $\rm 10^{11} M_\odot$ (solid line) and $\rm 10^{12} M_\odot$ (dash-dotted line) initial infall masses.}%
 \label{fig: abundances}%
\end{center}
\end{figure*}

\begin{figure*}
\begin{center}
 \subfloat{\includegraphics[width=1\columnwidth]{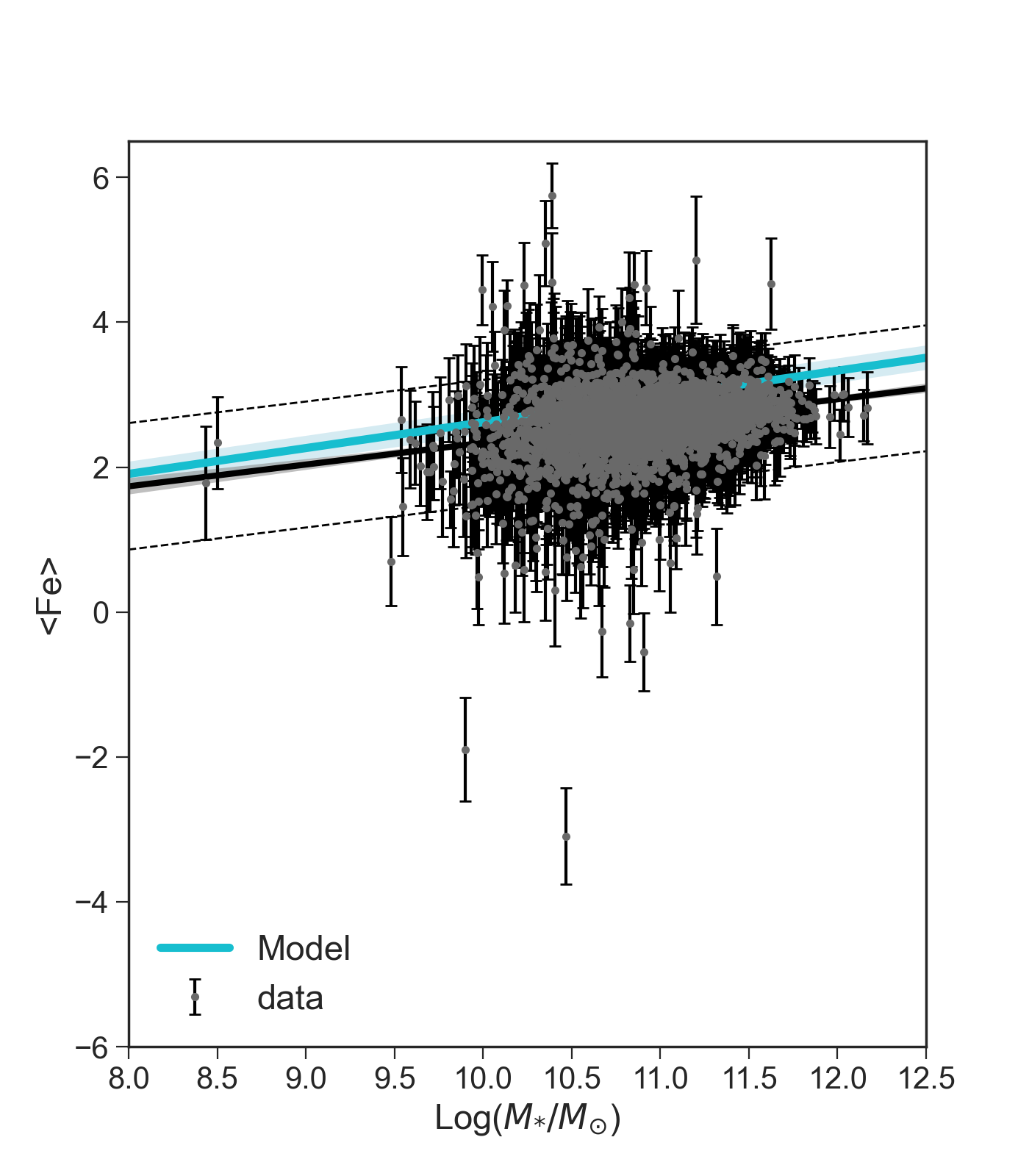}}
 \hfill
 \subfloat{\includegraphics[width=1\columnwidth]{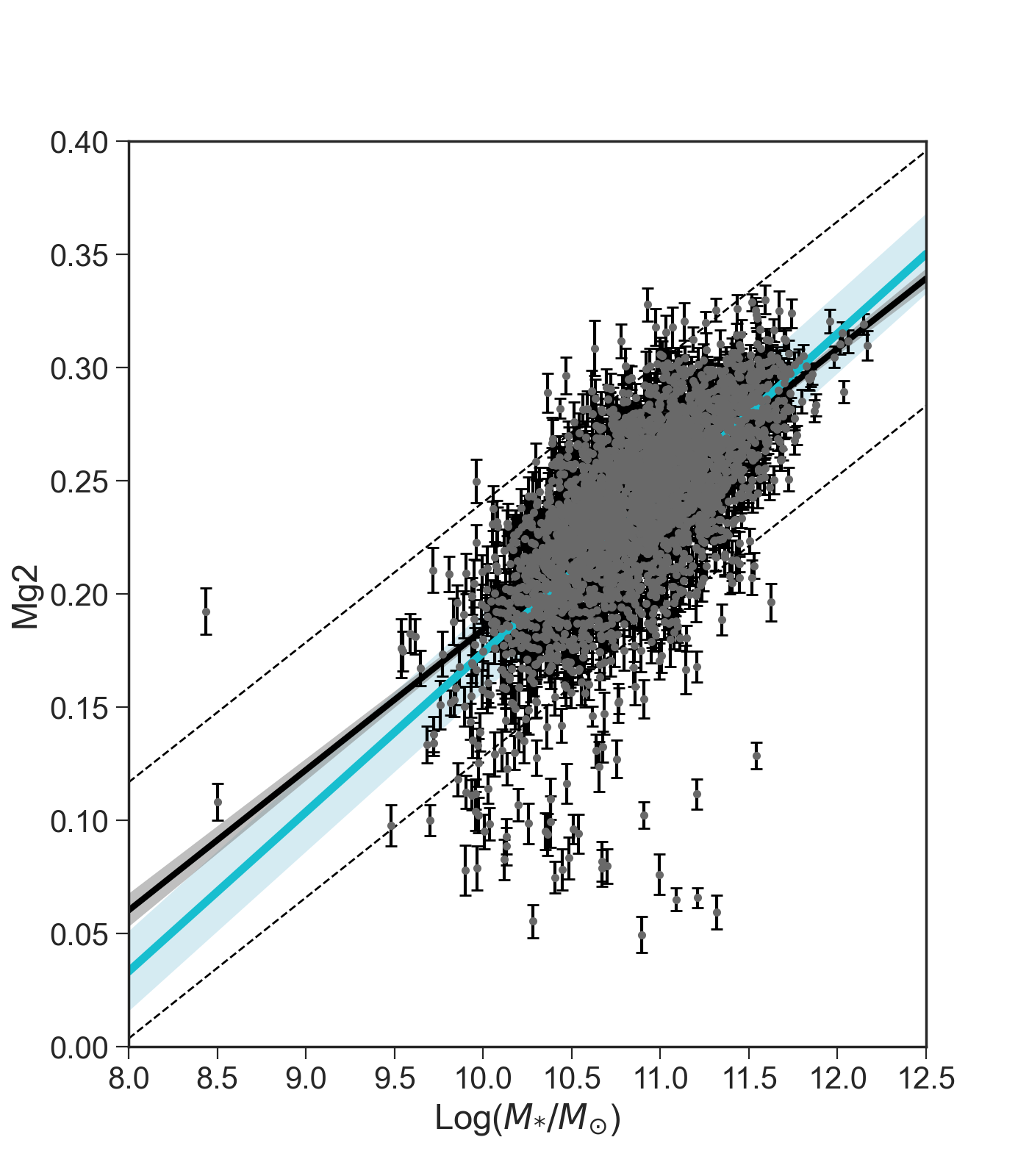}}%
 \caption{Line-strength indices predicted by the model using \citet{tantalo1998} calibrations together with observational data for both $\mathrm{<Fe>}$ (left panel) and $\mathrm{Mg_2}$ (right panel). Black dots are the galaxies in the catalogue and the lines are the linear fit to the data (black line) and to the model (cyan line). The shaded area represents the 1$\sigma$ uncertainties, while the black dotted lines are the boundaries of the $\rm 95\%$ confident region.}%
 \label{fig: Mg2}%
\end{center}
\end{figure*}


\subsection{[\texorpdfstring{$\alpha$/}FFe] ratio and mass-metallicity relations}
\label{sec: mass-metallicity relations}

Our chemical evolution code provides the evolution as a function of time of the abundances of chemical elements in the ISM. For instance, Figure \ref{fig: abundances} shows the abundances of different $\alpha$-elements for an elliptical galaxy of initial infall mass of $\rm 10^{11}\ M_\odot$ (left panel) and the [O/Fe] ratios for elliptical galaxies of different infall masses (right panel). As it is possible to see, we infer a higher [O/Fe] in the ISM of the more massive galaxies at fixed [Fe/H], as a consequence of the effect of the more efficient SFR in the brightest galaxies relative to the smaller ones. The metallicity of ellipticals is measured only by means of metallicity indices obtained from their integrated spectra. The most common metallicity indicators are $\rm Mg_2$ and $\rm <Fe>$. In order to pass from metallicity indices to [Fe/H] (and viceversa) one needs to adopt a suitable calibration. In order to compare the results of our models with the observed averaged stellar abundances of the dominant stellar populations in the galaxies in the dataset, we first need to compute the mean stellar abundance of the element X. This is defined by \citet{pagel1975} as:
\begin{equation}
    <X/H>\ \equiv\ <Z_X>\ =\ \frac{1}{S_0}\int_0^{S_0}Z_X(S)dS,
\end{equation}
where $\mathrm{S_0}$ is the total mass of stars ever born contributing to light at the present time. We recall that the right procedure should be that of averaging on the stellar luminosity at the present time since the observed indices are weighted on V-band luminosity (e.g. \citealp{arimoto1987, matteucci1998}). However, as it has been shown by \citet{matteucci1998}, results obtained by averaging on luminosity are not significantly different from those obtained by averaging on mass, at least for massive galaxies (see also \citealp{pipino2004, demasi2018}). Therefore, in this work we will refer only to mass-averaged metallicites. Once the mass-averaged abundances have been determined, we can convert them into spectral indices. This is done by using the calibration relations derived from \citet{tantalo1998}, who consider the Mg/Fe ratios:
\begin{equation}
\begin{cases}
    \ Mg_2 = 0.233 + 0.217<Mg/Fe>+\\
    \ \ \ \ \ \ \ \ \ \ \ \ \ \ \ \ +\ (0.153+0.120<Mg/Fe>)<Fe/H>\\
    \ <Fe> = 3.078 + 0.341<Mg/Fe>+\\
    \ \ \ \ \ \ \ \ \ \ \ \ \ \ \ \ +\ (1.654-0.307<Mg/Fe>)<Fe/H>.
\end{cases}
\end{equation}
In Figure \ref{fig: Mg2}, we compare the predictions of our model with the observational data. In particular, the continuous black and cyan lines are the linear regression of the data points and of the model results, respectively, with the shaded area representing the 1$\sigma$ uncertainties. The black dotted lines are the boundaries of the 95\% confident region. Our model fit reasonably well the observed mass-metallicity relation. In particular, the increasing trend of both $\mathrm{<Fe>}$ and $\mathrm{Mg_2}$ is successfully reproduced, although the predicted metallicity, especially at high masses, is slightly to high, reflecting in a higher $\rm Mg_2$ than the observed one. This difference could be due to different assumptions, such as the adopted IMF, the prescriptions for the yields, the adopted calibration or a combination of those factors. For the $\mathrm{<Fe>}$ we predict a slope of $\rm m_{<Fe>}^{model}=0.356\pm0.084$ to be compared to that of the best fitting line of the observational data equal to $\rm m_{<Fe>}^{data}0.301\pm0.019$, while for the $\rm Mg_2$ we predict a slope of $\rm m_{Mg_2}^{model}=0.070\pm0.035$ to be compared to $\rm m_{Mg_2}^{data}=0.062\pm0.001$.
Therefore, we have shown that our models can well reproduce the chemical properties of ETGs stellar populations, formed before the time at which the galactic wind occurs. At this point, we want to study the passive evolution of ETGs, after the main wind and consequent stop of SF, and the effects of SNIa and AGN feedback.

\begin{figure*}
    \centering
    \includegraphics[width=2.1\columnwidth]{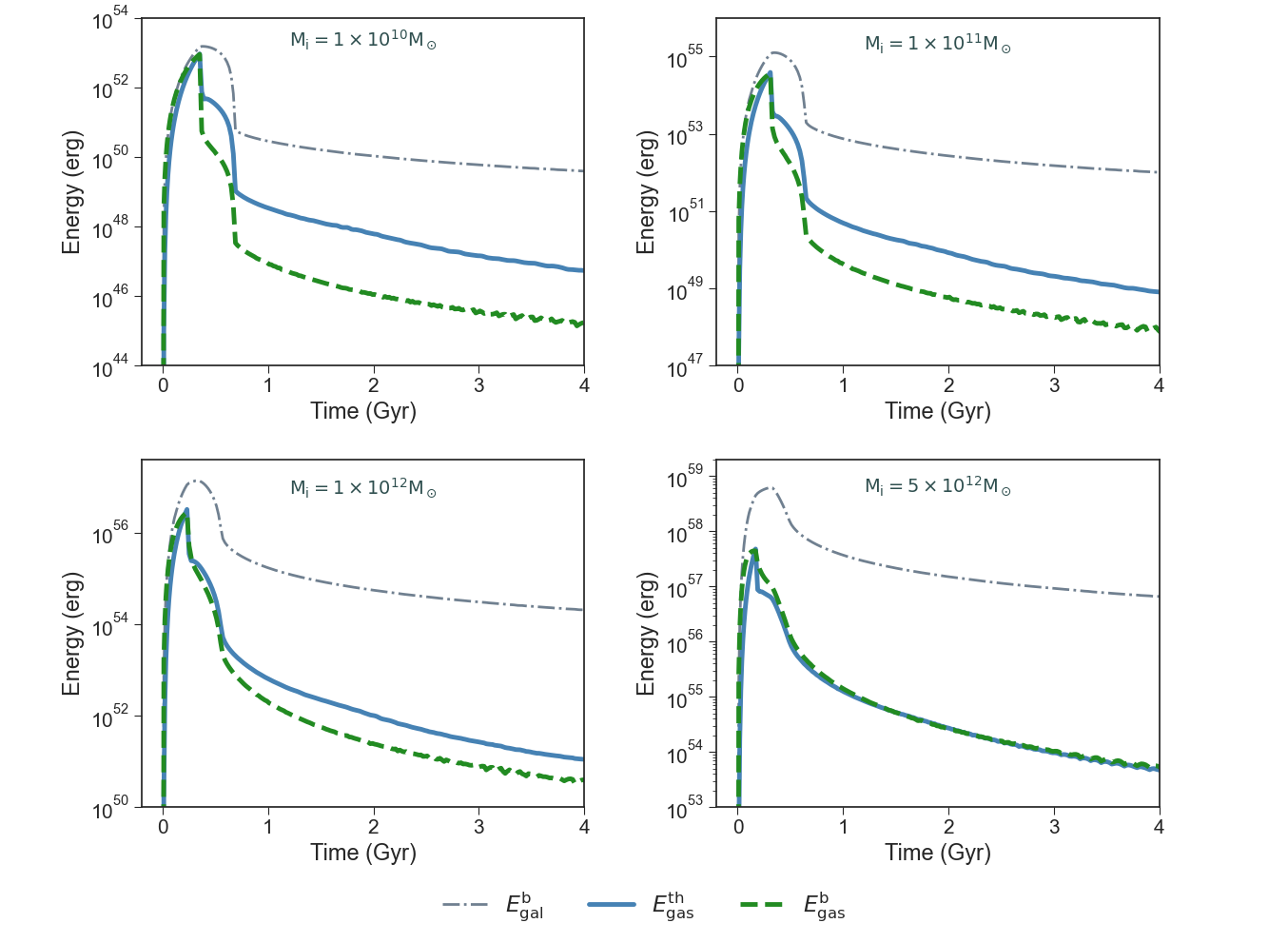}
    \caption{Energy balance as a function of time for simulated elliptical galaxies with initial infall mass of $\mathrm{10^{10}\ M_\odot}$, $\mathrm{10^{11}\ M_\odot}$, $\mathrm{10^{12}\ M_\odot}$ and $5\times\mathrm{10^{12}\ M_\odot}$. For each panel, the blue solid line represents the  thermal energy of the gas $\rm E_{gas}^{th}$, the green dashed line its binding energy, $\rm E_{gas}^{b}$, and the grey dash-dotted line the binding energy of the galaxy, $\rm E_{gal}^{b}$.}
    \label{fig: energetics}
\end{figure*}

\begin{figure*}
    \centering
    \includegraphics[width=2.1\columnwidth]{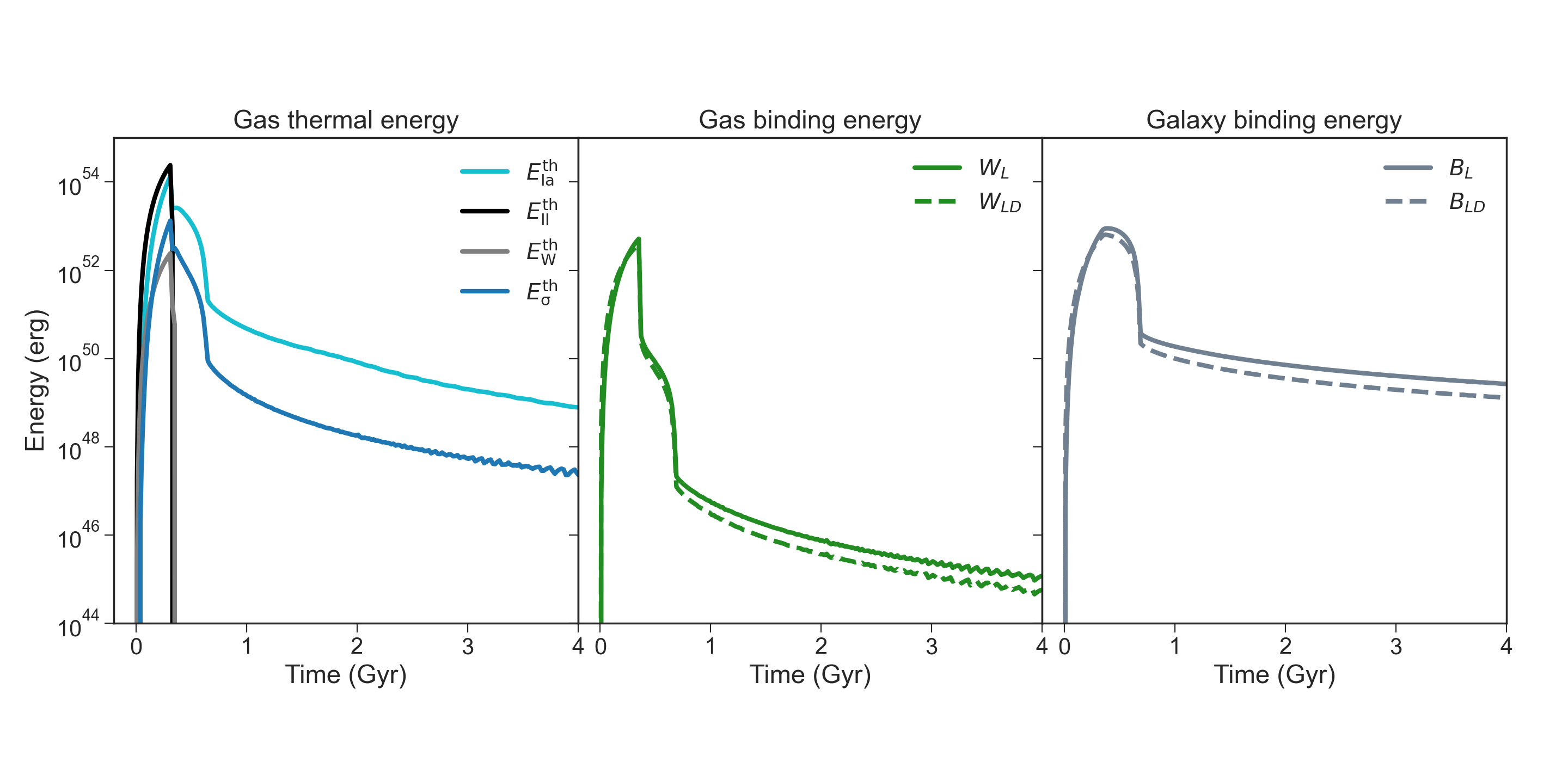}
    \caption{Energies for an elliptical galaxies with initial infall mass of $\mathrm{10^{11}\ M_\odot}$. Left panel: contribution to the gas thermal energy by Type II SNe ($\rm E_{II}^{th}$), Type Ia SNe ($\rm E_{Ia}^{th}$) and stellar wind ($\rm E_{W}^{th}$, $\rm E_{\sigma}^{th}$). Central panel: contribution to the total gas binding energy from the gravitational energy of the gas due to luminous matter, $\rm W_L$, and the gravitational energy of the gas due to the interaction between luminous and dark matter, $\rm W_{LD}$. Right panel: same as central panel, but for the galaxy binding energy.}
    \label{fig: components}
\end{figure*}

\begin{figure*}
    \centering
    \includegraphics[width=2.1\columnwidth]{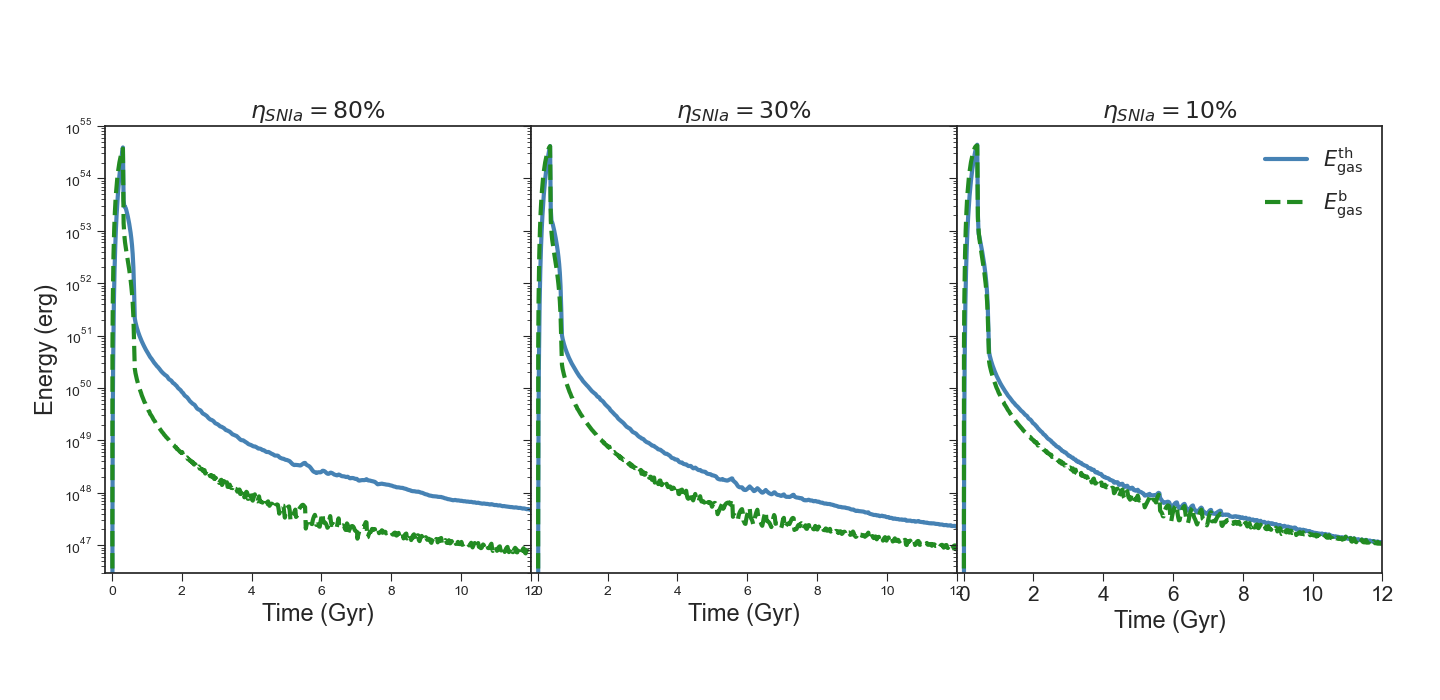}
    \caption{Comparison between the thermal energy of the gas, $\rm E_{gas}^{th}$ (blue solid line), and its binding energy, $\rm E_{gas}^{b}$ (green dotted line), for a galaxy of initial mass $\mathrm{M_i=10^{11}\ M_\odot}$, for different values of the efficiency of energy transfer from Type Ia SNe.}
    \label{fig: effsnia}
\end{figure*}

\subsection{Energies with no AGN feedback}
We start showing the results of models where the AGN feedback is not considered.
In Figure \ref{fig: energetics} we show the evolution as a function of time of the gas thermal energy ($\mathrm{E_{gas}^{th}}$), the gas binding energy ($\mathrm{E_{gas}^{b}}$) and the galaxy binding energy ($\mathrm{E_{gal}^{b}}$) for elliptical galaxies of initial infall mass of $\mathrm{10^{10}\ M_\odot}$, $\mathrm{10^{11}\ M_\odot}$, $\mathrm{10^{12}\ M_\odot}$ and $5\times\mathrm{10^{12}\ M_\odot}$, as predicted by the model.  

As explained in section \ref{sec: energetics}, when the thermal energy of the gas heated by SN explosions and stellar winds exceeds its binding energy, the gas present in the galaxy is swept away and the subsequent evolution of the system is determined only by the amount of matter and energy which is restored to the ISM by the dying stellar generations, namely low mass stars, and among SNe, only Type Ia SNe. Therefore, a fundamental point in the evolution of elliptical galaxies is the time of the onset of the galactic wind, $\mathrm{t_w}$. In particular, in each panel of Figure \ref{fig: energetics} it is possible to see the different values of $\mathrm{t_w}$, which coincide to the points at which the gas thermal energy becomes larger than the gas binding energy. With increasing galaxy mass, the value of $\mathrm{t_w}$ becomes smaller (as also reported in Table \ref{tab: models}) according to the \textit{inverse wind scenario} (\citealp{matteucci1994}). 

In order for a galaxy to be devoided of gas even after the time of the onset of the galactic wind,  the condition $\mathrm{E_{gas}^{th}\geq E_{gas}^{b}}$ must hold until the present time. For galaxies with mass $\rm M\le 10^{11}M_{\odot}$, this condition is easily reached. However, for systems of larger mass, and in particular for a galaxy of initial infall mass of $\mathrm{5\times10^{12}\ M_\odot}$ (corresponding to a final stellar mass of $\mathrm{1.5\times10^{12}\ M_\odot}$), the thermal energy of the gas appears to be comparable to its binding energy for all the evolution of the galaxy, creating a border line situation for the occurrence of a wind. 

In Figure \ref{fig: components} we report the evolution as a function of time of the different components of the total energy budget for a galaxy of initial infall mass of $\mathrm{10^{11}\ M_\odot}$. As discussed in section \ref{sec: energetics}, the ISM is heated by the thermalization of stellar motions ($\mathrm{E_{W}^{th}}$ and $\mathrm{E_{\sigma}^{th}}$) and SNe explosions (both Type II, $\mathrm{E_{II}^{th}}$, and Type Ia, $\mathrm{E_{Ia}^{th}}$). The contribution from Type II SNe dominates at early times but, as soon as the galactic wind occurs, it stops together with the contribution from stellar wind from massive stars. In fact, star formation halts when the thermal energy of the gas exceeds its binding energy. After the star formation has stopped, the galactic wind is maintained only by Type Ia SNe, which continue to explode until present time, and by the motion of lower mass stars. With a thermal energy of almost 2 orders of magnitude higher than that of stellar winds, Type Ia SNe appears to be the main drivers of the evolution of $\rm E_{gas}^{th}$, after the quenching of the SF. Here we assume an efficiency of energy transfer from Type Ia SNe equal to $\mathrm{\eta_{SNIa}=80\%}$ and justify our assumption by the fact that, since Type Ia SNe explosions occur in a medium already heated by Type II SNe, they should contribute to the total amount of their energy budget with minimal radiative losses (see \citealp{recchi2001}). However, we tested also other cases in which $\mathrm{\eta_{SNIa}}=30\%$ and $\mathrm{\eta_{SNIa}}=10\%$, whose results are reported in Figure \ref{fig: effsnia}, for a galaxy of initial infall mass equal to $\mathrm{10^{11}\ M_\odot}$. As one can see, when the efficiency of energy transfer of Type Ia SNe gets as low as $10\%$, the thermal energy of the gas appears to be almost comparable to its binding energy for all the galaxy evolution, so that basically the situation is the same of that illustrated previously in the lower right panel of Figure \ref{fig: energetics} for a high-mass system. 

It appears then reasonable to conclude that when no AGN feedback is considered, the thermal energy injected by SNe in the ISM is capable to both drive galactic winds at early times and to keep the inefficiency of the SF during the subsequent galaxy evolution. This is true at least for systems with $\rm M_i \leq10^{12}\ M_\odot$, but characterized by a high Type Ia SNe efficiency of energy transfer, or for systems of $\rm M_i \leq10^{11}\ M_\odot$ but characterized by a low efficiency of energy transfer. Therefore, in the following sections we will focus first on describing the treatment adopted to characterize the BH accretion and the AGN feedback, and then we will show its impact on the evolution and on the energy balance of a high-mass galaxy of $\rm 5\times10^{12}\ M_\odot$.


\section{BLACK HOLE ACCRETION AND AGN FEEDBACK}
\label{sec: black hole accretion and agn feedback}

In our phenomenological treatment of the AGN feedback we consider only radiative feedback, thus neglecting other mechanisms as radiation pressure and relativistic particles, as well as mechanical phenomena associated with jets. It is usually assumed that SMBHs are assembled by mergers with other BHs and/or by accretion of the gas from the surrounding medium. Theoretical studies suggest that a seed BH with the mass in the range $\rm10^2-10^6\ M_\odot$ (\citealp{2011valiante}) can form either by rapid collapse of Pop III stars (\citealp{2002heger}) or by the direct collapse of massive hot and dense gas clouds induced by gravitational instabilities (\citealp{2003bromm, 2006begelman, 2009volonteri}). In this work, we consider a BH of seed mass equal to $\rm 10^6 M_\odot$ which suffers spherical accretion of material at the Bondi rate (\citealp{bondi1952}):
\begin{equation}
    \dot{M}_B(t)= 4\pi R_B^2\rho_gc_s\lambda,
    \label{eq: Bondi}
\end{equation}
where $\rm \rho_g$ and $\rm c_s$ are the density and sound speed of the gas, respectively, and $\rm R_B$ is the Bondi radius, namely the gravitational radius of influence of the BH, given by:
\begin{equation}
    R_B=\frac{GM_{BH}\mu m_p}{\gamma k_b T_V},
\end{equation}
with $\rm \mu$ being the mean molecular weight of the gas, $\rm m_p$ the mass of the proton, $\rm k_b$ the Boltzmann constant, $\rm G$ the gravitational constant and $\rm \gamma$ the polytropic index ($\rm \gamma=1$ in the isothermal case). The parameter $\rm \lambda$ in equation \ref{eq: Bondi} is the dimensionless accretion parameter which, as determined by \citet{Ciotti2018} (see also \citealp{2022mancino}), can assume a wide range of values depending on the galaxy structure. Here we set $\rm \lambda=2\times10^4$ for a galaxy of initial infall mass $\rm M_i = 5\times10^{12}\ M_\odot$, even if this choice has a little  effect, since the BH growth will be Eddington limited during the entire period of interest. The virial temperature, $\rm T_V$, is given by:
\begin{equation}
    T_V=\frac{1}{3k_B}\mu m_p\sigma^2,
\end{equation}
with $\rm \sigma^2$ being the stellar velocity dispersion (see Section \ref{sec: thermal energy}). 

The accretion is limited to the Eddington rate, namely the accretion rate beyond which radiation pressure overwhelms gravity:
\begin{equation}
    \dot{M}_{Edd}(t) = \frac{L_{Edd}}{\eta c^2},
\end{equation}
where $\eta$ gives the mass to energy conversion efficiency. In this study we adopt a fixed value of $\eta=0.1$, which is the mean value for a radiatively efficient \citet{shakura} accretion onto a Schwarzschild BH, ignoring the possibility of radiatively inefficient accretion phases.

The accretion onto the BH is then 
\begin{equation}
    \dot{M}_{BH}(t)= 
    \begin{cases}
    \ \dot{M}_B(t)\ \ \ if\ \ \ \dot{M}_B(t) \leq \dot{M}_{Edd}(t)\\
    \ 10^{-3}\dot{M}_{Edd}(t)\ \ \ if\ \ \ \dot{M}_B(t) > \dot{M}_{Edd}(t),
    \end{cases}
\label{eq: accretion}
\end{equation}
where $\mathrm{\dot{M}_{B}(t)}$ is the rate from eq. \ref{eq: Bondi}. The corresponding bolometric luminosity is computed as:
\begin{equation}
    L_{BH}=\epsilon\dot{M}_{BH}c^2,
\end{equation}
where $\epsilon=0.1$. As it is possible to see from eq. \ref{eq: accretion}, we are using a reduction factor of $\mathrm{10^{-3}}$ to limit the maximum accretion rate. In an ideal simulation, the BH accretion at the Eddington rate limit would fluctuate in time, with shorter and shorter time scales at increasing spatial and temporal resolution, since the feedback time scale would decrease by moving nearer and nearer to the BH. Here, we are using a one-zone model, with a time step limited to $\mathrm{20\ Myr}$. Therefore, in our simulation, in absence of the reduction factor we would actually greatly overestimate the accretion. For what concerns the order of magnitude of the reduction factor, we explored different values in the range $\mathrm{10^{-3}-1}$. As expected, for the value equal to unity, i.e. for an unphysical continuous Eddington accretion lasting $\mathrm{20\ Myr}$, we found unrealistic results for both the BH accretion and, as a consequence, the BH mass $\mathrm{M_{BH}}$ and luminosity $\mathrm{L_{BH}}$. Similar results are obtained also for a reduction factor of $\mathrm{\sim 10^{-1}}$. Physically reliable solutions are obtained for a reduction factor in the range $\mathrm{10^{-3}-10^{-2}}$. We then chose the value of $\mathrm{10^{-3}}$ and based our consideration on \citet{2017ciotti} (see also \citealp{2007ciotti}) where a duty-cycle of the order of $\mathrm{10^{-3}}$ is commonly measured. In practice, the reduction factor should not be intended as a reduction of the feedback at the peak values of the AGN luminosity, but as a time-average over the length of the numerical time-step to be adopted in our one-zone simulations.


Finally, we compute the energy per unit time deposited by the BH into the ISM as:
\begin{equation}
    \dot{E}_{th}^{AGN}(t)=\xi L_{BH}T_V\bar{n}_p,
\end{equation}
with $\rm T_V$ expressed in $\rm K$ and $\rm \bar{n}_p$ being the average number of particles per $\rm cm^3$ near the galactic centre. We call the quantity $\xi$ the total absorption coefficient: following \citealp{2001ciotti} (equations 4 and A10) this parameter can be estimated in the optical thin regime with values of the order of $\sim 3\times10^{-14}$ for realistic galaxy sizes and ISM properties. For example, when $\rm \xi = 3\times10^{-14}$, $\rm T_V=10^7\ K$ and $\rm \bar{n}_p=10^2\ cm^{-3}$, only $\rm \sim 3\times10^{-5}$ is actually deposited as thermal energy in the ISM (see also \citet{1995binneytabor}). Of course, this number can change significantly during the galaxy evolution. Therefore, due to the intrinsic and unavoidable uncertainties on the value of $\xi$, in this work we test four different values, namely: $3\times10^{-14}$, $3\times10^{-4}$, $3\times10^{-2}$ and $1$, with this latter two being completely unphysical as they would certainly predict an AGN thermal feedback with an energy deposition larger than the available one. In the simulations we used also these extreme values in order to be sure that we bracketed the true behaviour.

\begin{figure*}
    \centering
    \includegraphics[width=2.1\columnwidth]{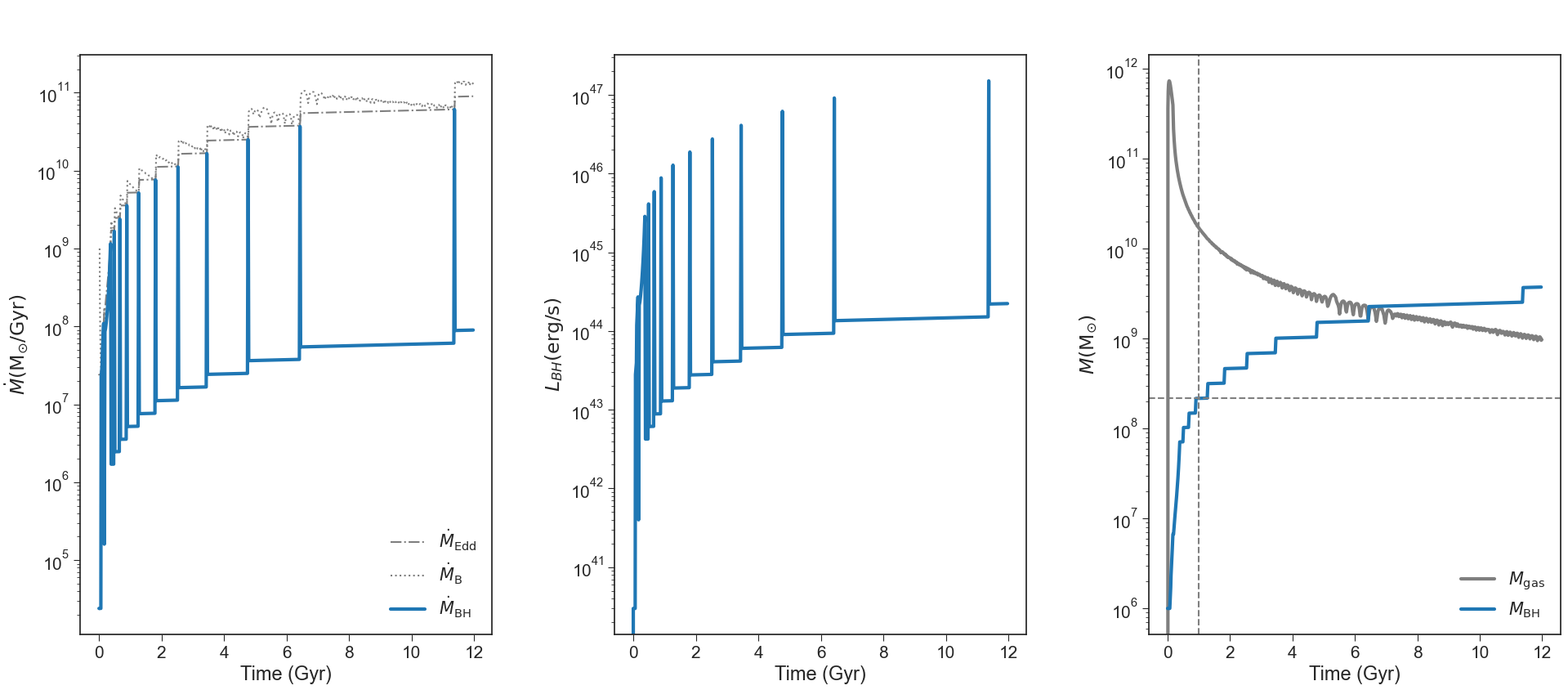}
    \caption{Evolution of the accretion rate, bolometric luminosity and BH mass for an elliptical galaxy of initial infall mass $\rm 5x10^{12} M_\odot$. Left panel: evolution of the accretion rate as a function of time. In grey dotted line is reported the Bondi accretion rate, in grey dash-dotted line is reported the Eddington accretion rate and in cyan continuous line the resulting accretion according to equation \ref{eq: accretion}. Central panel: bolometric luminosity evolution as a function of time. Right panel: the BH mass evolution as a function of time, with the vertical grey dotted lines indicating the mass reached by the BH after $\rm 1\ Gyr$. The grey continuous line represent the evolution of the mass of gas inside the galaxy.}
    \label{fig: accretion}
\end{figure*}

\begin{figure*}
\begin{center}
 \subfloat{\includegraphics[width=1\columnwidth]{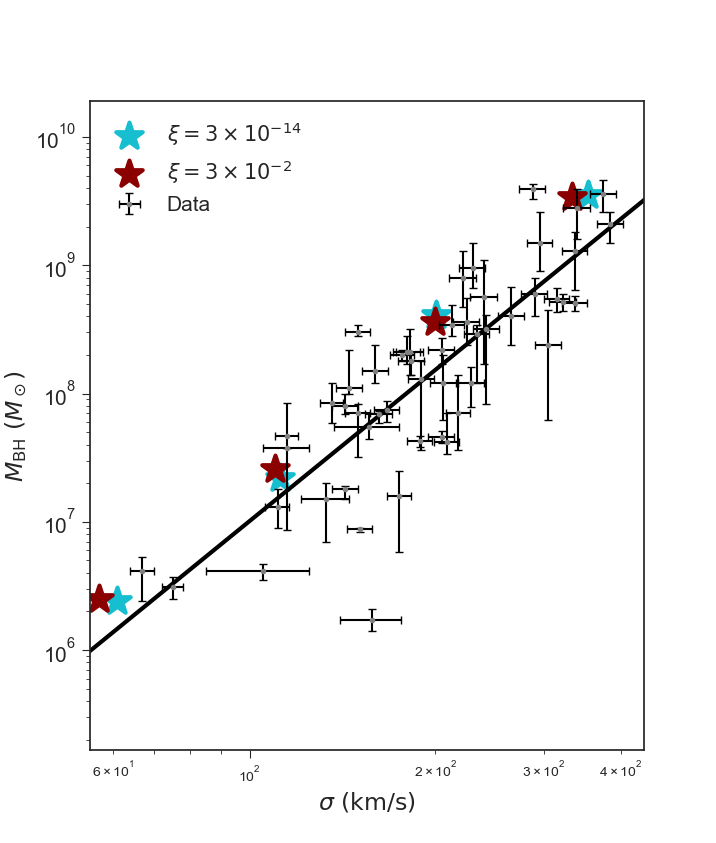}\label{fig:a}}
 \hfill
 \subfloat{\includegraphics[width=1\columnwidth]{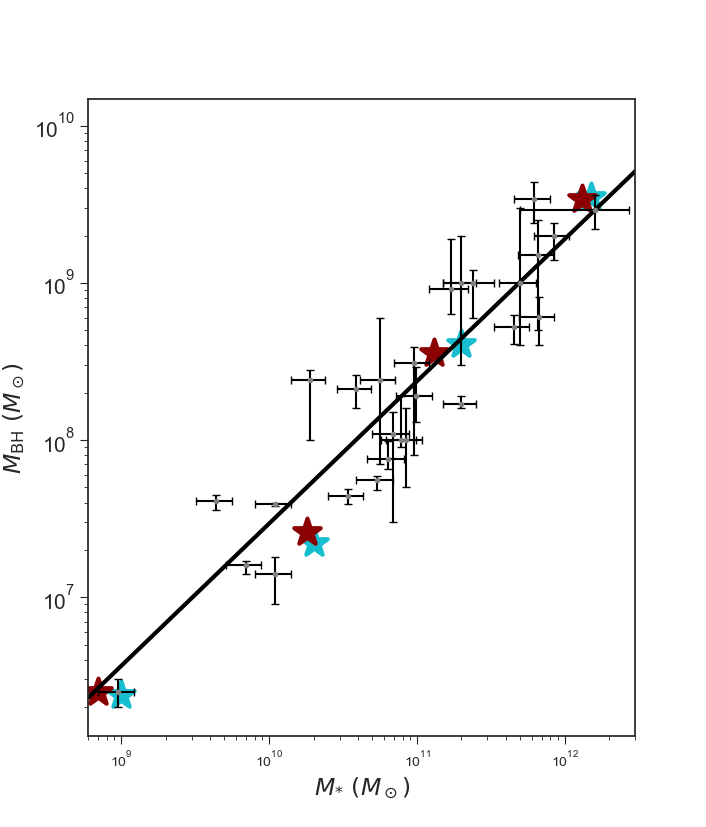}\label{fig:b}}%
 \caption{Comparison between our model predictions and observational data from \citet{2010gultekin} for the $\rm M_{BH}$ vs $\rm \sigma$ relation (left panel) and for the $\rm M_{BH}$ vs $\rm M_{*}$ relation with observational data from \citet{2003marconihunt} (right panel). In both panel, the solid black line is the best fit of the observational data and the cyan and red stars are the predictions of our model for galaxies of initial infall mass equal to $\rm 10^{10}$, $\rm 10^{11}$, $\rm 10^{12}$ and $\rm 5\times10^{12}\ M_\odot$ and for an absorption coefficient $\xi$ equal to $\mathrm{3\times10^{-14}}$ and $\mathrm{3\times10^{-2}}$, respectively.}%
 \label{fig: scatter}%
\end{center}
\end{figure*}

\subsection{Black hole masses and luminosities}

In Figure \ref{fig: accretion} we show the accretion onto the BH evolution as a function of time together with the corresponding bolometric luminosity and BH mass evolution (central and right panel, respectively). As it is possible to see from Figure \ref{fig: accretion}, the resulting accretion rate evolution is characterized by a series of spikes, each with a duration of $\rm 40 Myr$, corresponding to the moments at which $ \dot{\rm M}_{\rm B}\leq\dot{\rm M}_{\rm E}$. The spikes are reflected into the luminosity, being this latter proportional to $\rm \dot{M}_{BH}$ (see central panel of the same Figure) and, as a consequence, it is characterized by a burst shape representative of the highly intermittent activity that QSOs may exhibit. The predicted final value of the luminosity is $\rm L_{BH}=2.2\times10^{44}\ erg/s$ which is three order of magnitudes lower than the value that it assumes in the last burst, equal to $\rm L_{BH}=1.43\times10^{47}\ erg/s$. It must be noted that with these values we find a very good agreement with several observations of AGN bolometric luminosity both at high and at lower redshift (\citealp{2010dunn, 2017fiore, 2021izumidue, 2021izumiuno}).

The BH reaches a mass of $\rm 2\times10^8\ M_\odot$ after 1 Gyr of galaxy evolution and a final mass of $\rm 3.5\times10^{9}\ M_\odot$ at the present time. It is well known that there exist well-defined correlations between the mass of the SMBH, $\rm M_{BH}$, and the properties (e.g. velocity dispersion, $\rm \sigma$, and the stellar mass, $\rm M_{*}$) of the spheroidal component of the host galaxy (\citealp{1998magorrian}). Even if there have been claims for a non-linear relation between $\rm M_{BH}$ and $\rm M_{*}$ (\citealp{2001laor, 2001wuhan}), \citet{2003marconihunt} re-established the tight linear relation: $\rm <M_{BH}/M_{*}>\sim 0.002$, in good agreement also with several other estimates (e.g.: \citealp{2002mcluredunlop, 2003dunlop, 2004haring}). In Figure \ref{fig: scatter} we compare predictions of our model for galaxies of initial infall mass in the $\rm 10^{10}-5\times10^{12}\ M_\odot$ range with estimates for the $\rm M_{BH}$ vs $\sigma$ (by \citealp{2010gultekin}) and for the $\rm M_{BH}$ vs $\rm M_{*}$ (by \citealp{2003marconihunt}) relations. We remind $\sigma^2=0.335GM_*(t)/R_e$ being the stellar velocity dispersion (see Section \ref{sec: thermal energy}). The predictions we are showing here have been obtained with the physical expected value of $\rm \xi = 3\times10^{-14}$ and show good agreement with observations. We show also what happen when higher values of $\rm \xi$ ($\mathrm{3\times10^{-2}}$) are adopted. The results change very little, without affecting the agreement between measurements and predictions. This is a noticeable result, given the simplicity of our model.

\begin{figure*}
    \centering
    \includegraphics[width=2.1\columnwidth]{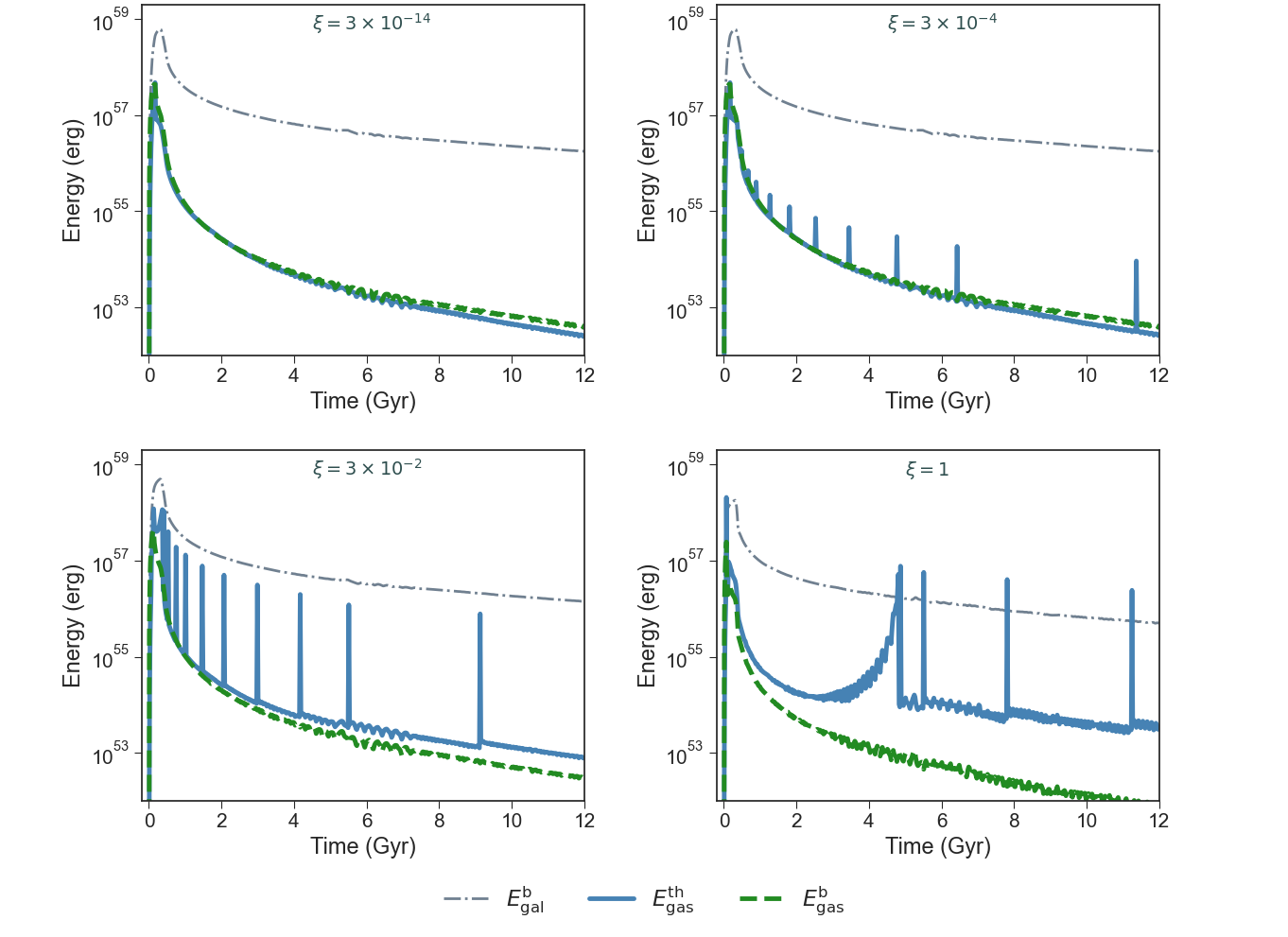}
    \caption{Energy balance as a function of time for simulated elliptical galaxies with initial infall mass of $5\times\mathrm{10^{12}\ M_\odot}$ and for different values of the absorption coefficient $\xi$. For each panel, the blue solid line represents the  thermal energy of the gas $\rm E_{gas}^{th}$, the green dashed line its binding energy, $\rm E_{gas}^{b}$, and the grey dash-dotted line the binding energy of the galaxy, $\rm E_{gal}^{b}$.}
    \label{fig: energies_AGN}
\end{figure*}

\begin{figure*}
    \centering
    \includegraphics[width=2.1\columnwidth]{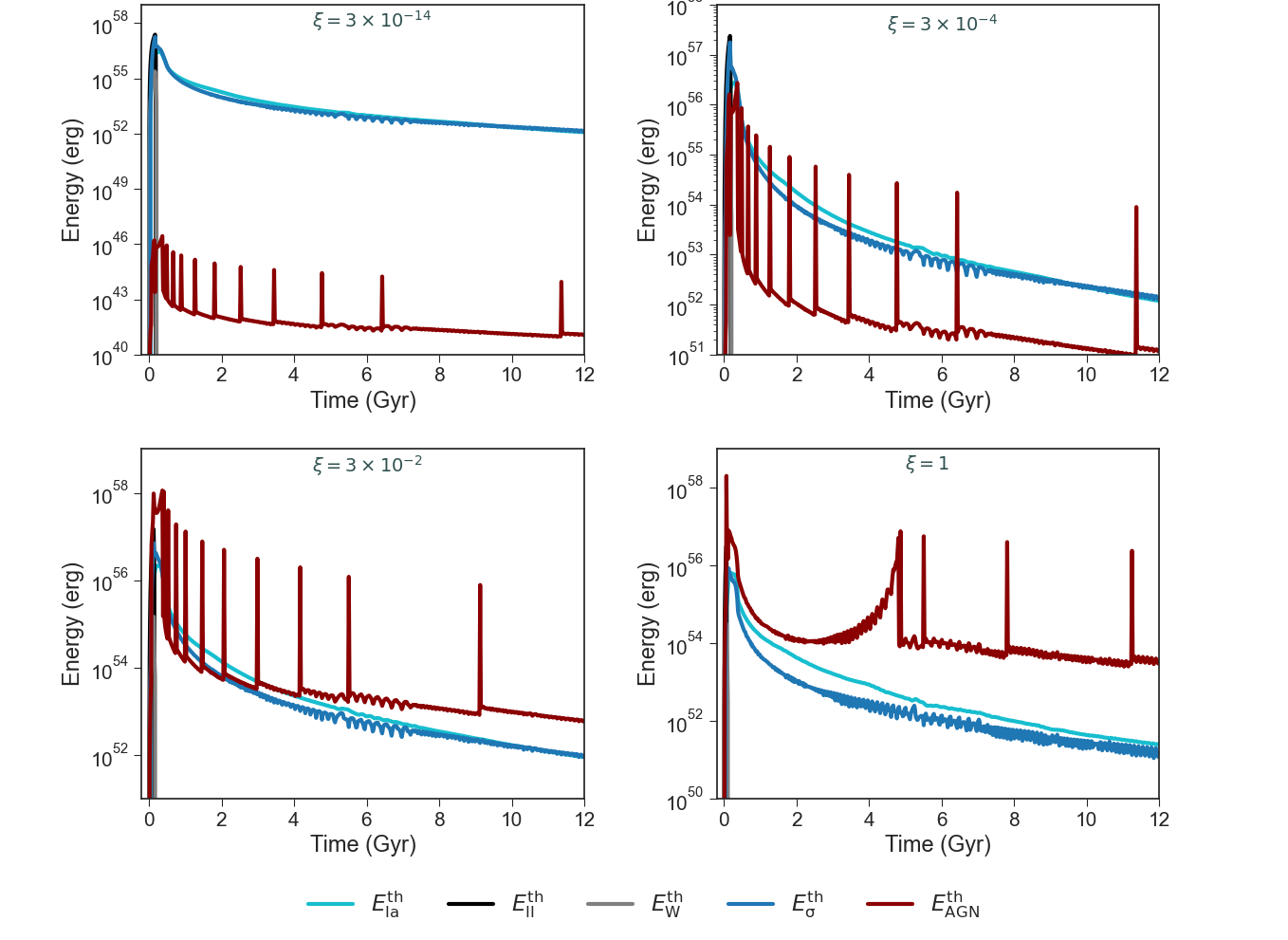}
    \caption{Contributions to the gas thermal energy by Type II SNe ($\rm E_{II}^{th}$), Type Ia SNe ($\rm E_{Ia}^{th}$), stellar wind ($\rm E_{W}^{th}$, $\rm E_{\sigma}^{th}$) and AGN feedback ($\rm E_{AGN}^{th}$), for different values of the absorption coefficient $\xi$.}
    \label{fig: components_AGN}
\end{figure*}


\begin{figure}
    \centering
    \includegraphics[width=1\columnwidth]{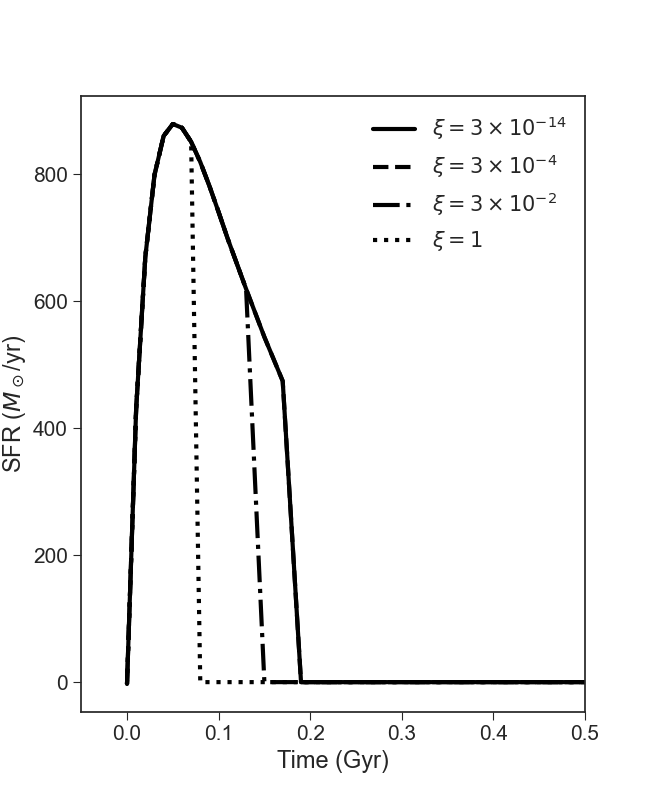}
    \caption{Evolution of the SFR for a galaxy with initial infall mass of $\rm 5\times10^{12}\ M_\odot$ and for different values of the absorption coefficient.}
    \label{fig: SFR}
\end{figure}

\subsection{Energies with AGN feedback}

Figure \ref{fig: energies_AGN} shows the evolution as a function of time of the gas thermal energy ($\mathrm{E_{gas}^{th}}$), the gas binding energy ($\mathrm{E_{gas}^{b}}$) and the galaxy binding energy ($\mathrm{E_{gal}^{b}}$) for an elliptical galaxy of initial infall mass of $\rm 5\times10^{12}M_\odot$ and for four different values of the coefficient $\xi$ ($3\times10^{-14}$, $3\times10^{-4}$, $3\times10^{-2}$, $1$).

In the upper left panel of Figure \ref{fig: energies_AGN} are reported results for models with $\xi=3\times10^{-14}$. For this model, the effect of the AGN in the evolution of the thermal energy of the gas is totally negligible and there is no difference between this situation and that in which AGN feedback is not considered. The coefficient $\xi$ must be increased by at least ten orders of magnitude before the effect of the AGN on the thermal energy of the gas appears to be visible (upper left panel of Figure \ref{fig: energies_AGN}). Even if the contribution from the AGN appears to be no longer negligible, the time at which the galactic wind starts is equal to that of the models with $\xi=3\times10^{-14}$ and with no AGN feedback ($\rm t_{GW}=0.19\ Gyr$). Therefore, the AGN feedback cannot be the main cause of the formation of a galactic wind. However, since its contribution to the subsequent evolution of $\rm E_{gas}^{th}(t)$ cannot be neglected, its role could be crucial in maintaining quenched the galaxy after SF suppression. The situation drastically changes if one adopts a coefficient as high as $\xi=3\times10^{-2}$. In this case, the total thermal energy of the gas, $\rm E_{gas}^{th}(t)$, becomes completely dominated by the AGN feedback for all its evolution. In this model, the galactic wind starts also at earlier times, $\rm t_{GW}=0.15\ Gyr$, so that the AGN feedback seems to be its main driver. Finally, in the lower right panel of Figure \ref{fig: energies_AGN} we show the results for the energy evolution in the case in which $\xi=1$. As one can see, the many bursts that characterize the shape of $\rm E_{gas}^{th}$, are so powerful to exceed the binding energy of the galaxy $\rm E_{gal}^b$. Since the physical consequence of this is a BH which could potentially disrupt  entirely the host galaxy or at the very least remove a very large fraction of its gas, it seems physically unreasonable that the AGN feedback process could be so efficient.

The situation is illustrated in more details in the upper left panel of Figure \ref{fig: components_AGN}, where we show the evolution of the different components of the thermal energy. When $\xi=3\times10^{-14}$, the thermal energy due to the AGN is several orders of magnitude lower than that due to the other phenomena. In particular, the thermalization is dominated by the contribution of stellar motions and SN explosions, both at early and at late times (with Type II SNe being the major contributors at early times and Type Ia SNe and stellar motions at late times). The evolution of $\rm E_{AGN}^{th}(t)$ reflects the bursty accretion history of the BH, as expected. Due to the high accretion episodes that we described in the previous section, the BH injects powerful bursts of thermal energy into the surrounding gas. However, the coefficient $\xi$ must be increased of at least ten orders of magnitude before the effect of the AGN on the thermal energy of the gas appears to be visible. In fact, in this case the bursts in the AGN thermal energy appears to be comparable with the energy injected by SNe and stars. Even if the contribution from the AGN is no longer negligible, at early times the thermal energy evolution is still dominated by Type II SNe which continue to be the main driver of the galactic winds. For a coefficient $\xi=3\times10^{-2}$ the thermal energy injected into the ISM by the AGN is larger also than that due to Type II SNe and this causes the AGN feedback to be the main driver of the galactic wind. Finally, for the model with $\xi=1$, the energy injected by the AGN is much larger than that due to the other phenomena and, as stated above, this cause an unrealistic evolution of the considered galaxy. 

In Figure \ref{fig: SFR} we report the evolution of the SFR as a function of time for the four different cases corresponding to the different values of $\xi$. As it is possible to see, the halt of the SF happens at the same time in the model with $\xi=1\times10^{-14}$ and in the model with $\xi=1\times10^{-4}$, as well as in the model in which we do not consider the effect of the AGN feedback. On the other hand, for the models with $\xi=1\times10^{-2}$ and $\xi=1$, the SF stops at $\rm 150\ Myr$ and $\rm 80\ Myr$, respectively, due to the non negligible affect of the AGN feedback at earlier time.

\section{Conclusions}
  
In this work, we modelled the evolution of ETGs with initial infall mass in the range $\rm 10^{10}-5\times10^{12}\ M_\odot$ by means of a chemical evolution model able to follow the evolution with time of the gas mass and its chemical composition during the entire galactic lifetime. In this first paper, we focused on the effects of stellar and AGN feedback and their role in suppressing the SF in ellipticals at early times. In order to do that, we updated the computation of the energetic budget in our model which now includes, besides core-collapse and Type Ia SNe, both stellar winds from LIMS and AGN feedback. In this way, the ISM is heated by stellar winds (both from massive stars and LIMS), SNe of all types and AGN feedback, and whenever its thermal energy exceeds its binding energy a strong galactic wind is generated and the SF is suppressed. We recall that as far as SMBH accretion (and consequently AGN feedback) is concerned, we also take into account the effect of radiation pressure which stops accretion when the luminosity exceeds the Eddington luminosity. Therefore, even if not directly, radiation pressure influences the ISM thermal energy. We paid particular attention to the role of Type Ia SNe feedback in the suppression of SF and the maintenance of such situation, after the main galactic wind.
 
\subsection{Models without AGN feedback}

In the first set of simulations presented in this paper we excluded the contribution of AGN feedback from the energetic budget. The thermal energy of the gas depends on SNe (both Type Ia and core-collapse), and on stellar winds (both from massive stars and LIMS). After the occurrence of the first galactic wind and consequent suppression of SF, only Type Ia SNe contribute to the thermal energy of the gas which is restored by low mass stars, with the additional contribution of the thermalization due to the velocity dispersion of the ejecta from the dying low mass stars.
We first tested the model without AGN feedback on the chemical properties of the dominant stellar populations in ellipticals (e.g. mass metallicity relation and [$\alpha$/Fe] ratios) and selected the parameters that best reproduce observations. From the point of view of the energetic budget, the main conclusions are the following:

\begin{itemize}
    \item By assuming an efficiency of energy transfer of $\eta_{II}=3\%$ and of $\eta_{Ia}=80\%$ for core-collapse and Type Ia SNe (see \citealp{recchi2001}), respectively, all systems are able to develop a first massive galactic wind, when the condition $\rm E_{gas}^{th}\geq E_{gas}^{b}$ is satisfied. The time of the onset of the galactic wind, $\rm t_{gw}$, becomes smaller at increasing galaxy mass, according to the \textit{inverse wind scenario} of \citet{matteucci1994}, and this is due to an assumed increasing efficiency of star formation with galactic mass.
    
    \item  All systems with final stellar mass $\rm \lesssim 10^{12}\ M_\odot$ can satisfy the condition $\rm E_{gas}^{th}\geq E_{gas}^{b}$, for the entire galaxy life, when the above SN efficiencies of energy transfer are adopted. In other words, these galaxies are suffering a continuous wind for the remaining $\rm \sim 12 Gyr$, after the main early wind. However, for higher mass systems the thermal energy of the gas, after the main wind, appears to be comparable to its binding energy for all the passive period of the evolution of the galaxy, thus creating a situation in which the gas is not lost from the system.
    
    \item If instead the efficiency of energy transfer of Type Ia SNe is assumed to be as low as $\eta_{Ia}=10\%$, the situation of comparable thermal and binding energy of the gas, after the main wind, occurs for systems of lower stellar mass ($\rm \sim 10^{10}\ M_\odot$), but for all the smaller galaxies persists the situation of a continuous wind triggered mainly by Type Ia SNe.
\end{itemize}

Therefore, it appears reasonable to conclude that when  AGN feedback is not considered, the thermal energy injected by Type Ia and core-collapse SNe in the ISM is enough for driving global galactic winds at early times as well as to keep the SF quenched for the entire period of passive evolution. In particular, the SF is quenched either in systems with stellar mass $\rm \lesssim 10^{12}\ M_\odot$, but characterized by a high Type Ia SNe efficiency of energy transfer ($\rm \sim 80\%$), or in systems with stellar mass $\rm \lesssim 10^{10}\ M_\odot$, but with an efficiency of energy transfer as low as ($\rm \sim 10\%$). As a consequence, it appears that for high mass galaxies an additional source of energy should be required, in particular if the efficiency of energy transfer by Type Ia SNe is significantly smaller than $\rm \sim 80\%$, and this additional energy should be provided by the AGN feedback.

\subsection{Models with AGN feedback}

In our study we adopted as the additional source of heating the AGN feedback. We considered the effects of radiative feedback on a galaxy with initial infall mass of $\rm 5\times10^{12} M_\odot$ (corresponding to a final stellar mass of $\rm 1.5\times10^{12}\ M_\odot$), neglecting a direct effect of radiation pressure and/or other mechanisms associated with jets. Radiation pressure (as parametrized by the Eddington luminosity) plays an indirect role on ISM heating, because BH accretion and the associated energy injection are stopped whenever the accretion luminosity is larger than the Eddington one. In the simulation, the central BH is characterized by a seed mass of $\rm 10^6 M_\odot$ and, as just recalled, it undergoes standard Bondi-Eddington limited accretion. Due to the one-zone nature of our model, we are forced to fix an absorption coefficient $\xi$, namely the fraction of accretion luminosity actually deposited on the ISM via Compton heating. This number, in hydrodynamical simulations with radiative transport, is found to be time dependent. Here we use the parametrization introduced in \citet{1997ciotti}, but we change the value by orders of magnitude, exploring also some unrealistic cases. As the results change very little even for large variations in the adopted value of $\xi$, we are confident that our conclusions are quite robust. 
In particular, we considered four different values for the absorption coefficient $\xi$, i.e. $\rm 3\times10^{-14}$ (the physically expected order of magnitude for a realistic gaseous atmosphere of an elliptical galaxy, see Section \ref{sec: black hole accretion and agn feedback}), $\rm 3\times10^{-4}$, $\rm 3\times10^{-2}$ and $\rm 1$ (unrealistically high values used to test the importance of AGN thermal feedback), which allowed us to isolate four different physical situations. We reached the following conclusions:

\begin{itemize}
    \item As expected, due to the indirect role of the radiation pressure which reduces the BH accretion whenever the accreted luminosity is larger than the Eddington one, the evolution of the BH accretion rate is characterized by a series of spikes, each with a duration of $\rm \sim 40\ Myr$. The spikes are reflected into the luminosity which, as a consequence, is characterized by a bursting shape. The predicted bolometric luminosities are in the range $\rm 10^{44}-10^{47}\ erg/s$, in good agreement with observations.

    \item For absorption coefficients below $\rm \xi=3\times10^{-4}$, the effect of the AGN on the evolution of the thermal energy of the gas is totally negligible, with no difference between this model and the one without AGN feedback. 
    
    \item For $\rm \xi=3\times10^{-4}$ the effect of the AGN on the thermal energy of the gas becomes detectable, however the time at which the galactic wind starts is unchanged with respect to the model without AGN feedback. Therefore, in this scenario, the AGN cannot be the main cause for the formation of a galactic wind. However, since  its contribution for the subsequent evolution of $\rm E_{gas}^{th}$ cannot be neglected, its role can be crucial in maintaining the SF quenched.
    
    \item For $\rm \xi=3\times10^{-2}$, the total thermal energy of the gas becomes completely dominated by the AGN feedback during the entire evolution. In this model, the galactic wind also sets in at earlier times so that the AGN feedback appears to be its main driver together with core-collapse SNe.
    
    \item In the unphysical case in which $\rm \xi=1$, the many bursts that, due to the AGN feedback, characterize the shape of $\rm E_{gas}^{th}$, are so powerful that they can provide an energy exceeding the binding energy of the entire galaxy. Therefore, we are inclined to consider this case physically unacceptable. 
    
    \item We computed the final BH masses for galaxies of initial infall mass equal to $\rm 10^{10}$, $\rm10^{11}$, $\rm10^{12}$ and $\rm 5\times10^{12}\ M_\odot$. We succeeded in reproducing the observed proportionality between the stellar mass of the host galaxy and that of the central black hole as well as the Magorrian relation, without the need of stopping ad hoc the accretion.
\end{itemize}

In conclusion, the most convincingly scenario is the one in which the ISM is thermalized by both AGN feedback and SNe of all types. When the efficiency of energy transfer of Type Ia SNe is $\rm \sim 80\%$, core-collapse and Type Ia SNe are capable of both driving a global galactic wind at early times and at keeping the SF quenched during the passive evolution for systems with stellar mass $\rm \lesssim 10^{12}\ M_\odot$. If one adopts only an efficiency of $\rm \sim 10\%$  for Type Ia SNe, simulating the strong cooling present in the innermost galaxy regions, then the galaxy stellar mass above which AGN feedback is necessary is $\rm \sim 10^{10}\ M_\odot$. The cooling process is indeed a complex one and depends strongly on the environmental conditions. When SNe explode in a cold and dense medium, the cooling is quite effective. On the other hand, when the environment is warm and rarefied the cooling is negligible. For example \citet{recchi2001}, by means of a dynamical model, suggested that the feedback of Type Ia SNe is more effective than that from Type II SNe (\citealp{cioffi1988,bradamante1998,2004melioli}), since the former explode in an already heated medium. When the contribution from the AGN is added and is characterized by the physically expected value for the absorption coefficient of $\rm \xi=3\times10^{-14}$, the BH feedback appears to be important to regulate the growth of the BH itself but only marginally important for the galaxy evolution. The first effects on the thermalization of the ISM manifest when an absorption coefficient $\rm \xi \simeq 10^{-4}$ is adopted. In that case, the effect of the AGN on the development of the main galactic wind is still negligible when compared to that of SNe, but it can substantially contribute in keeping the SF quenched during the galaxy passive evolution. This result is supported also by recent hydrodynamical simulations. In particular, \citet{2021lanfranchi} (see also \citealp{2015caproni,2017caproni}) investigated the effects of outflows from BHs on the gas dynamics in dwarf spheroidal galaxies (dSphs) by means of 3D hydrodynamic simulations, and concluded that, in an inhomogeneous ISM, the impact of the AGN outflow appears to be substantially reduced and its contribution to the removal of gas from the galaxy is almost negligible.

\section*{Acknowledgements}
 We thank the anonymous referee for the useful comments and suggestions. 

\section*{Data Availability}

The data underlying this article will be shared upon request.



\bibliographystyle{mnras}
\bibliography{example} 








\bsp	
\label{lastpage}
\end{document}